\documentclass[10pt,conference]{IEEEtran}
\IEEEoverridecommandlockouts
\usepackage{cite}
\usepackage{amsmath,amssymb,amsfonts}
\usepackage{algorithmic}
\usepackage{graphicx}
\usepackage{textcomp}
\usepackage{xcolor}
\def\BibTeX{{\rm B\kern-.05em{\sc i\kern-.025em b}\kern-.08em
    T\kern-.1667em\lower.7ex\hbox{E}\kern-.125emX}}

\usepackage{hyperref}
\usepackage{url}            
\usepackage{booktabs}       
\usepackage{bm}
\usepackage{bbding}
\usepackage{multirow}
\usepackage{subfigure}
\usepackage{fancyhdr}

\graphicspath{ {images/} }

\begin{document}

\title{CertPri: Certifiable Prioritization for Deep Neural Networks via Movement Cost in Feature Space}


\author{
\IEEEauthorblockN{Haibin Zheng}
\IEEEauthorblockA{\textit{Zhejiang University of Technology} \\
Hangzhou, China \\
haibinzheng320@gmail.com}
\and
\IEEEauthorblockN{Jinyin Chen*}
\IEEEauthorblockA{\textit{Zhejiang University of Technology} \\
Hangzhou, China \\
chenjinyin@zjut.edu.cn}
\and
\IEEEauthorblockN{Haibo Jin}
\IEEEauthorblockA{\textit{Zhejiang University of Technology} \\
Hangzhou, China \\
2112003035@zjut.edu.cn}
\thanks{*Corresponding author.}
\thanks{This paper was accepted by the 38th IEEE/ACM International Conference on Automated Software Engineering (ASE 2023).}
}





\maketitle

\begin{abstract}
Deep neural networks (DNNs) have demonstrated their outperformance in various software systems,
but also exhibit misbehavior and even result in irreversible disasters.
Therefore, it is crucial to identify the misbehavior of DNN-based software and improve DNNs’ quality.
Test input prioritization is one of the most appealing ways to guarantee DNNs' quality,
which prioritizes test inputs so that more bug-revealing inputs can be identified earlier with limited
time and manual labeling efforts. 
However, 
the existing prioritization methods are still limited from three aspects: 
certifiability, effectiveness, and generalizability. 
To overcome the challenges, 
we propose \emph{CertPri}, 
a test input prioritization technique designed based on a movement cost perspective of test inputs in DNNs' feature space.
CertPri differs from previous works in three key aspects: 
(1)~\emph{certifiable} - it provides a formal robustness guarantee for the movement cost;
(2)~\emph{effective} - it leverages formally guaranteed movement costs to identify malicious bug-revealing inputs; and
(3)~\emph{generic} - it can be applied to various tasks, data, models, and scenarios.
Extensive evaluations across 
2 tasks (i.e., classification and regression),
6 data forms, 
4 model structures, 
and 2 scenarios (i.e., white-box and black-box)
demonstrate CertPri’s superior performance. 
For instance,
it significantly improves 53.97\% prioritization effectiveness on average compared with baselines.
Its robustness and generalizability are 1.41$\sim$2.00 times and 1.33$\sim$3.39 times that of baselines on average, respectively.
The code of CertPri is open-sourced at~\url{https://anonymous.4open.science/r/CertPri}.
\end{abstract}

\begin{IEEEkeywords}
Deep neural network, test input prioritization, deep learning testing, movement cost, certifiable prioritization
\end{IEEEkeywords}

\section{Introduction\label{Intro}}
Deep neural networks (DNNs)~\cite{LeCun2015DeepLearning} have performed impressive success in many fields, including
computer vision~\cite{Simonyan2015VGGNet,He2016ResNet,Alex2012AlexNet},
natural language processing~\cite{Alshemali2021NLP,Shuang2020NLP},
and software engineering~\cite{Zhang2019Apricot,Chen2020Software}, etc.
However, like traditional software systems,
DNNs are also vulnerable in terms of quality and reliability~\cite{Tian2018DeepTest,Eniser2019DeepFault,Hu2019DeepMutation++,Pei2017DeepXplore}.
Meanwhile, 
these vulnerabilities could lead to serious losses such as a crash caused by a Google self-driving car~\cite{Chris2016Google}, 
and even irreversible disasters such as the deadly car crash caused by the autopilots of Tesla~\cite{Jack2018Tesla} and Uber~\cite{Alex2018Uber}. 
Therefore, it is crucial to detect the misbehavior of DNN-based software and ensure DNNs' quality. 

Much effort has been put into DNN-based software quality assurance~\cite{Pei2017DeepXplore,Zheng2022NeuronFair,Fabrice2020IsNeuron,Xie2019DeepHunter,Guo2018Dlfuzz,Ma2018Deepgauge}.
One of the most appealing methods is test input prioritization~\cite{Chen2020PACE,Byun2019DNNsentiment},
which prioritizes test inputs so that more bug-revealing inputs 
(e.g., misclassified inputs) can be identified earlier with limited time and manual labeling efforts.
Moreover, 
these inputs facilitate the debugging of DNN-based software,
which could improve DNNs' quality~\cite{Feng2020DeepGini,Wang2021PRIMA} and reduce their retraining cost~\cite{Shen2020MCP}.
There are several prioritization methods,
mainly including four aspects, i.e.,
coverage-based~\cite{Ma2018Deepgauge,Pei2017DeepXplore,Wicker2018Feature},
surprise-based~\cite{Zhang2020NAF,Byun2019DNNsentiment,Kim2019Guiding},
confidence-based~\cite{Feng2020DeepGini,Shen2020MCP,Zhang2019NSA}, and
mutation-based~\cite{Wang2021PRIMA} methods.
The first two methods
prioritize test inputs based on DNNs' neuron coverage and surprise-adequacy activation traces, respectively.
Confidence-based methods identify bug-revealing inputs by measuring the classifier's output probabilities.
Mutation-based methods design a series of mutation operations, 
and then analyze the mutated output probabilities based on supervised learning.
These methods make great progress in identifying bug-revealing inputs earlier,
but they still suffer from the following problems.

First, 
the existing methods are empirically lacking formal guarantees, 
which results in vulnerability against malicious attacks,
i.e.,
prioritizing bug-revealing inputs at the back when attacked.
More specifically,
taking neuron activation suppression as an optimization objective, 
the adversary could craft imperceptible adversarial perturbations~\cite{Goodfellow2015FGSM,Dong2018MIFGSM}. 
These bug-revealing inputs with perturbations will be prioritized at the back due to low neuron coverage or inadequate activation traces,
i.e., coverage-based and surprise-based methods lose their effectiveness.
Similarly, confidence-based methods fail to work 
when increasing the highest probability value~\cite{Chen2020MAGGAN,Moosavi2016DeepFool} 
by adding well-designed perturbations to inputs.
Mutation-based methods include input mutation and model mutation, 
both of which are similar to data augmentation~\cite{Shorten2019DataAug,Sun2019STL} and network modification~\cite{Mustafa2019RHS,Rakin2019PNI}, respectively.
Once the mutation operations are leaked, 
the adversary can bypass these operations by crafting malicious bug-revealing inputs.
Then, mutation-based methods will be invalid.
Therefore, a formal robustness guarantee for certifiable prioritization is required.

Second, almost all methods either suffer from prioritization effectiveness or efficiency issues.
For instance,
coverage-based methods have been demonstrated to be ineffective and time-costly~\cite{Feng2020DeepGini}.
Surprise-based methods improve test input prioritization by utilizing more advanced metrics
(e.g., surprise-adequacy and activation frequency),
but are computationally expensive due to more parameter tuning.
Confidence-based methods apply output probabilities to perform fast and lightweight prioritization, 
and their effectiveness is better than the previous two. 
However, once adversarial~\cite{Chen2020MAGGAN} or poisoned~\cite{Li2021Deeppayload} inputs are injected into the test dataset, 
their effectiveness drops largely.
Mutation-based test input prioritization, PRIMA~\cite{Wang2021PRIMA}, 
is the state-of-the-art (SOTA) method for DNNs,
outperforming the confidence-based methods by an average of 10\%, 
but with a time cost increase of more than 100 times.
Moreover, PRIMA is a supervised prioritization method, 
i.e., its effectiveness is affected by training dataset size and category balance.

Third, most methods suffer from the generalizability issue,
including the generalizability of tasks, data forms, model structures, and application scenarios.
For instance,
confidence-based methods rely on DNNs' output probabilities, 
and thus may not be directly generalized to a regression task.
Meanwhile, 
their prioritization effectiveness for sequential data form (e.g., text data~\cite{Alshemali2021NLP}) has been demonstrated to drop by an average of 30\% in the existing study~\cite{Wang2021PRIMA}.
Mutation-based methods could be generalized to various data and tasks, 
but specialized domain knowledge is required to design diverse data-specific (e.g., structured data~\cite{Bellamy2018AIF360} and graph data~\cite{Zhang2020GraphsSurvey}) mutation strategies.
Moreover, 
it is unclear whether their model mutation strategies can still perform well on other model structures, 
such as graph convolution network (GCN)~\cite{Zhang2020GraphsSurvey}.
Except for confidence-based methods, 
the other three types are all designed for white-box testing and need to acquire DNNs' details. 
In black-box scenarios, 
these white-box methods will seriously degrade performance or even fail to work.

To overcome these challenges, 
our design goals are as follows: 
(1)~we intend to take formal guarantee into account when designing a certifiable prioritization method;
(2)~we want the certifiability to serve prioritization effectiveness without degrading efficiency;
(3)~we plan to evaluate its generalizability.

\begin{figure}[t]
    \centering
    \subfigure[\footnotesize(0,0.1)]{\includegraphics[width=0.19\linewidth,height=0.16\linewidth]{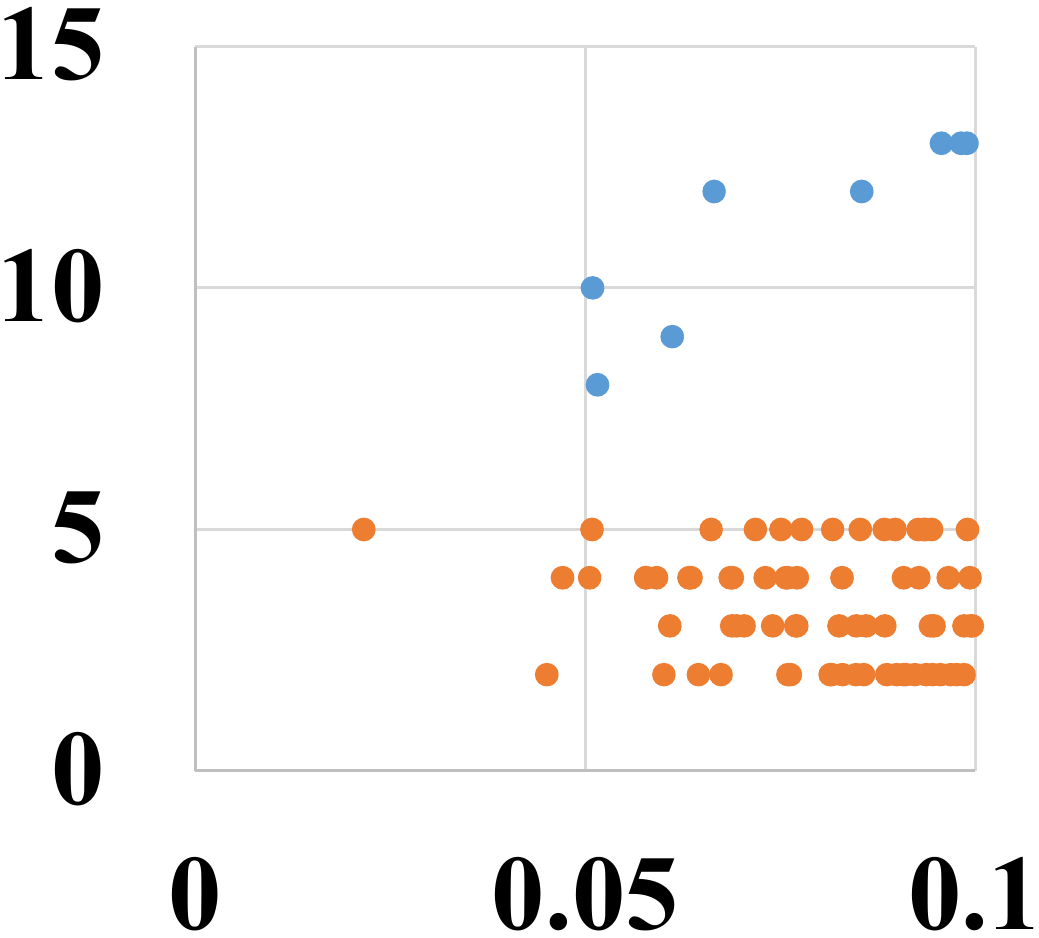}}
    \subfigure[\footnotesize[0.1,0.2)]{\includegraphics[width=0.19\linewidth,height=0.16\linewidth]{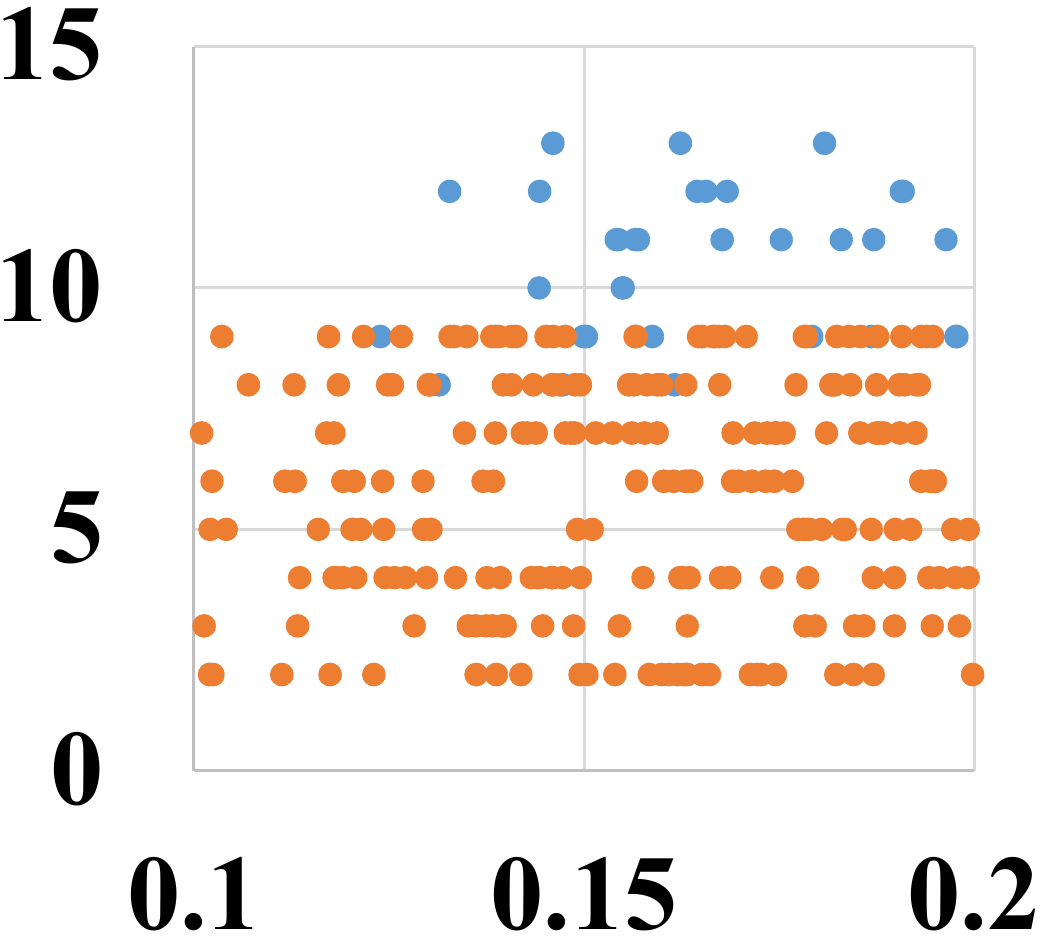}}
    \subfigure[\footnotesize[0.2,0.3)]{\includegraphics[width=0.19\linewidth,height=0.16\linewidth]{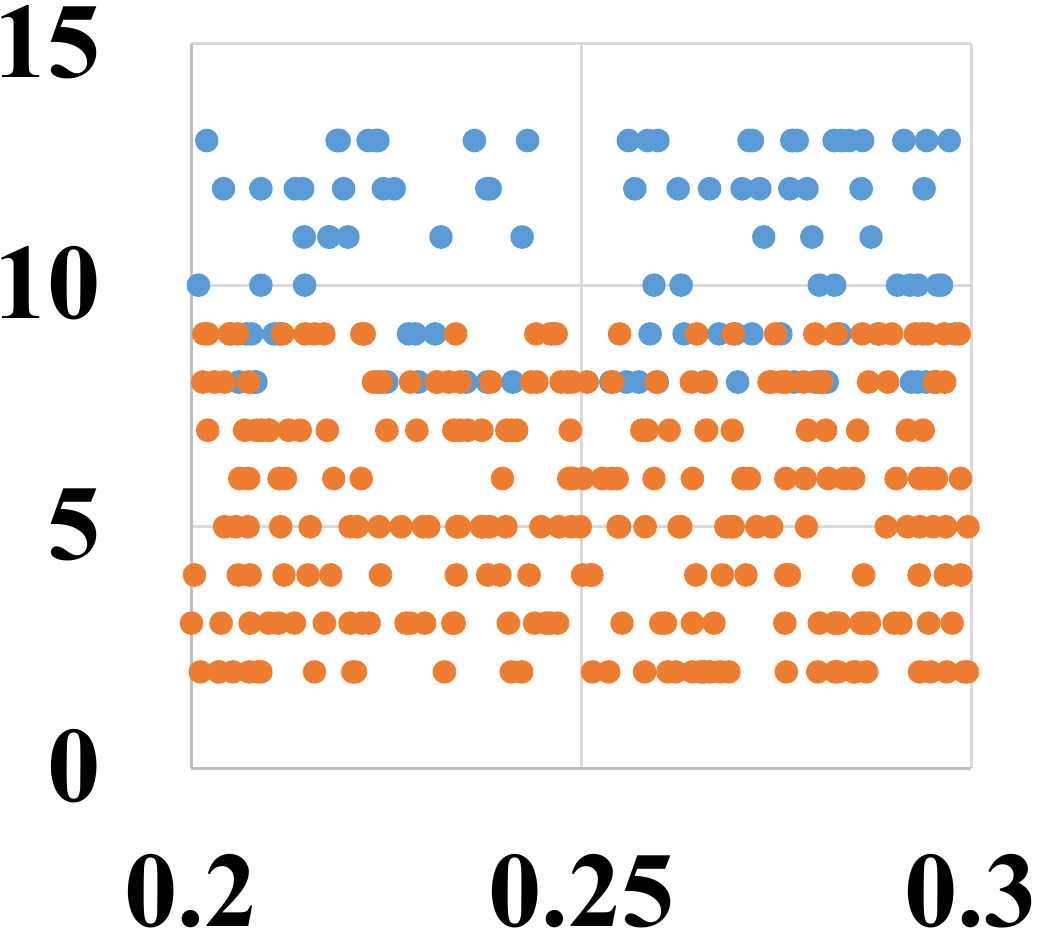}}
    \subfigure[\footnotesize[0.3,0.4)]{\includegraphics[width=0.19\linewidth,height=0.16\linewidth]{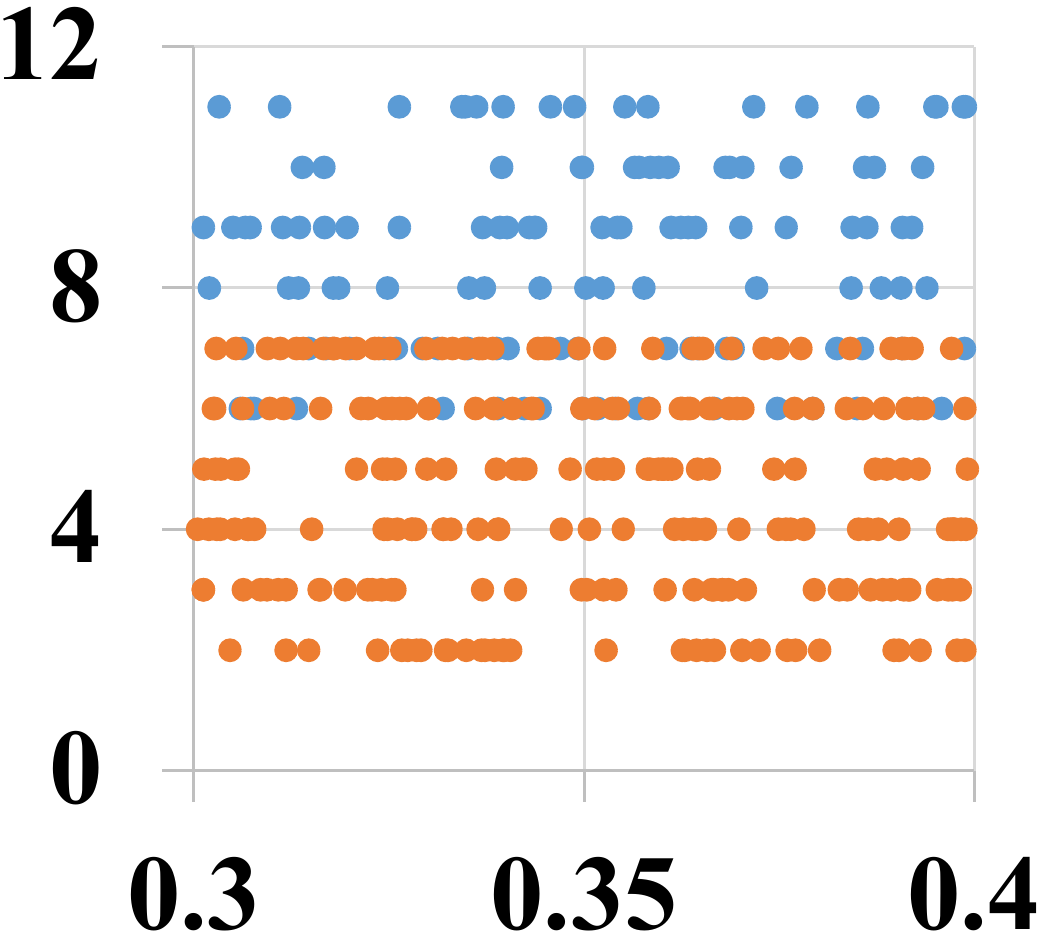}}
    \subfigure[\footnotesize[0.4,0.5)]{\includegraphics[width=0.19\linewidth,height=0.16\linewidth]{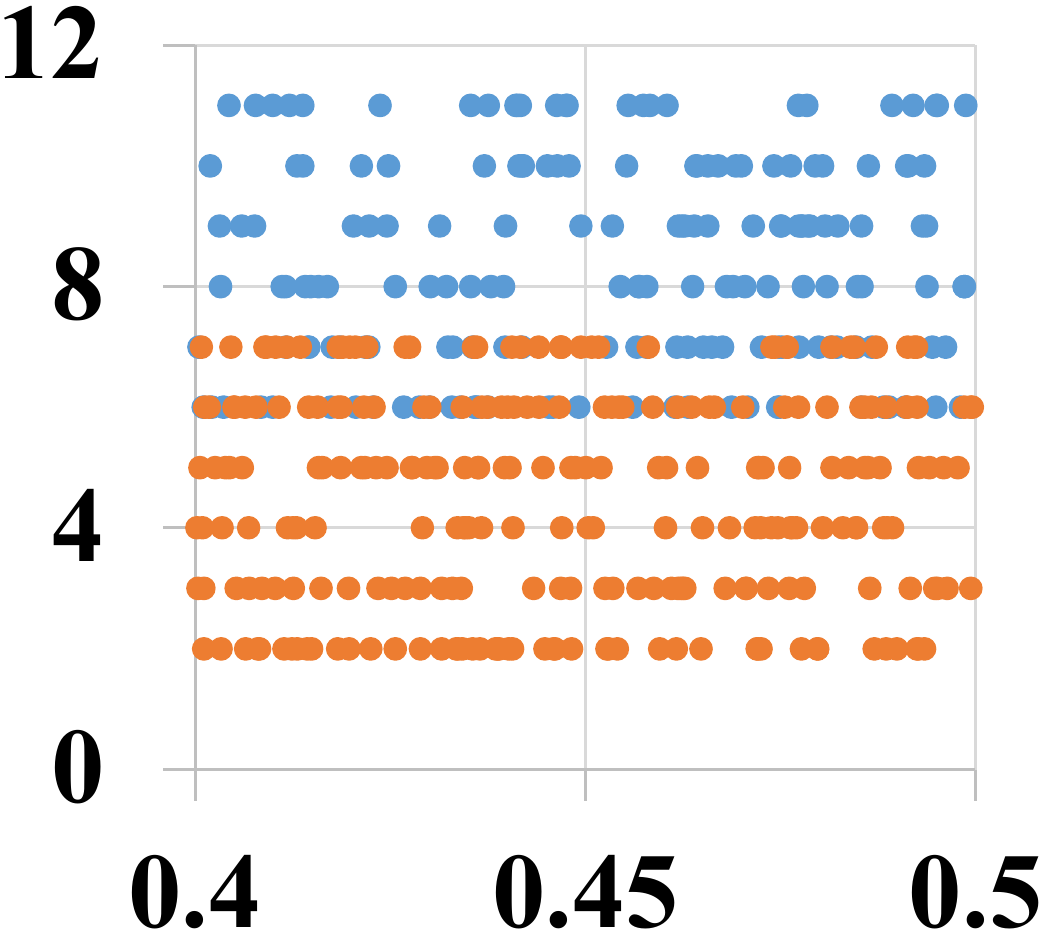}}
    \subfigure[\footnotesize[0.5,0.6)]{\includegraphics[width=0.19\linewidth,height=0.16\linewidth]{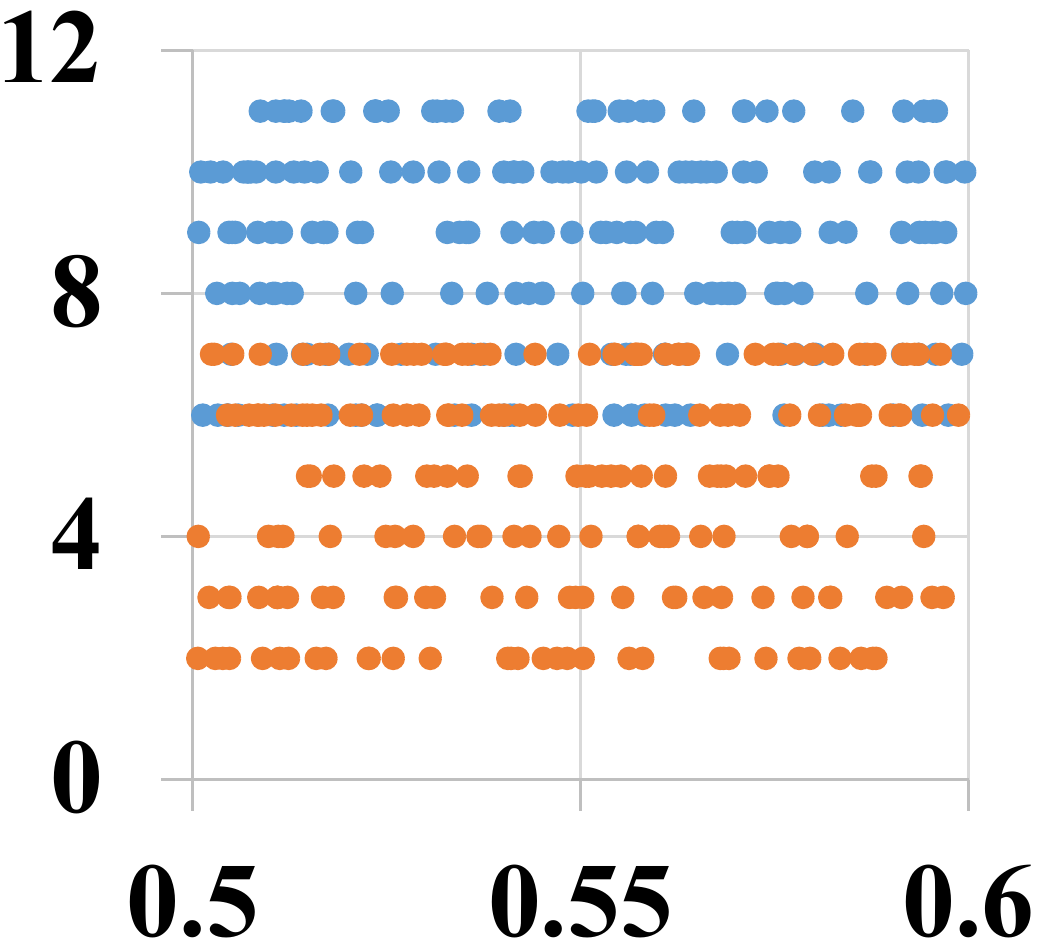}}
    \subfigure[\footnotesize[0.6,0.7)]{\includegraphics[width=0.19\linewidth,height=0.16\linewidth]{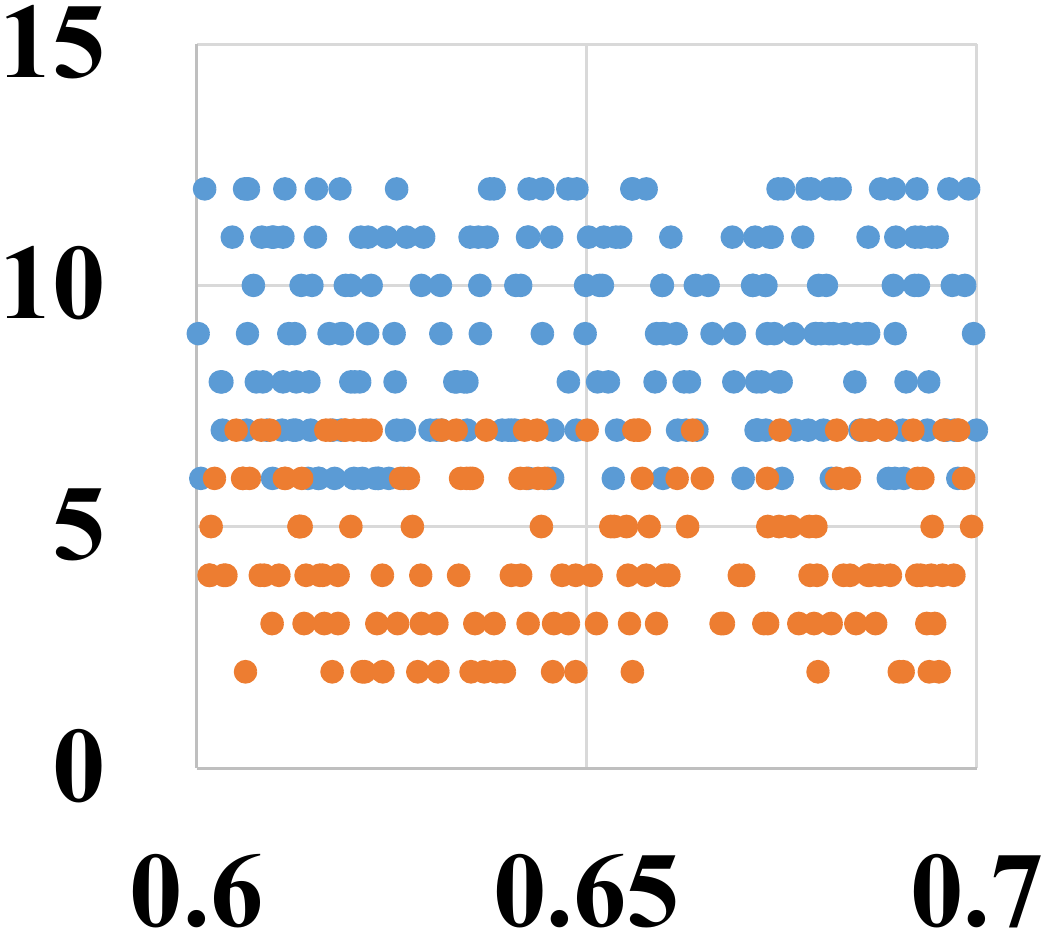}}
    \subfigure[\footnotesize[0.7,0.8)]{\includegraphics[width=0.19\linewidth,height=0.16\linewidth]{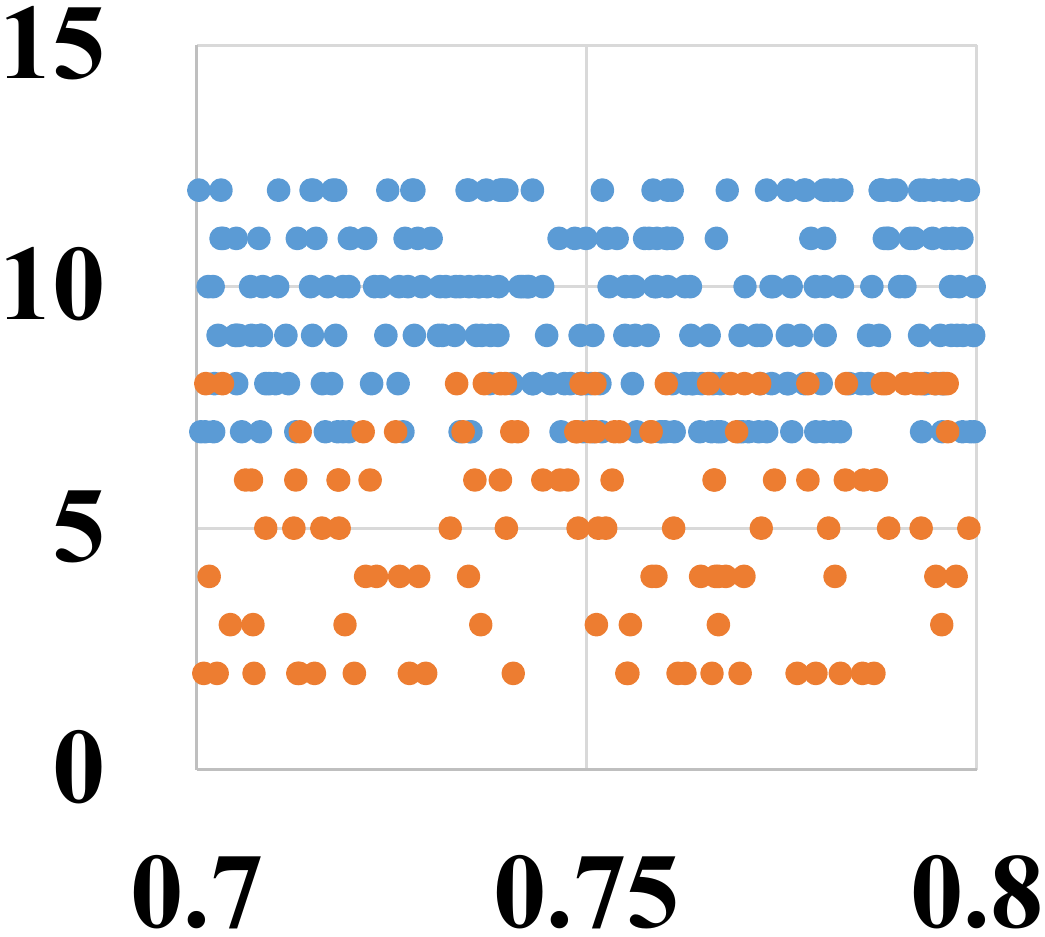}}
    \subfigure[\footnotesize[0.8,0.9)]{\includegraphics[width=0.18\linewidth,height=0.16\linewidth]{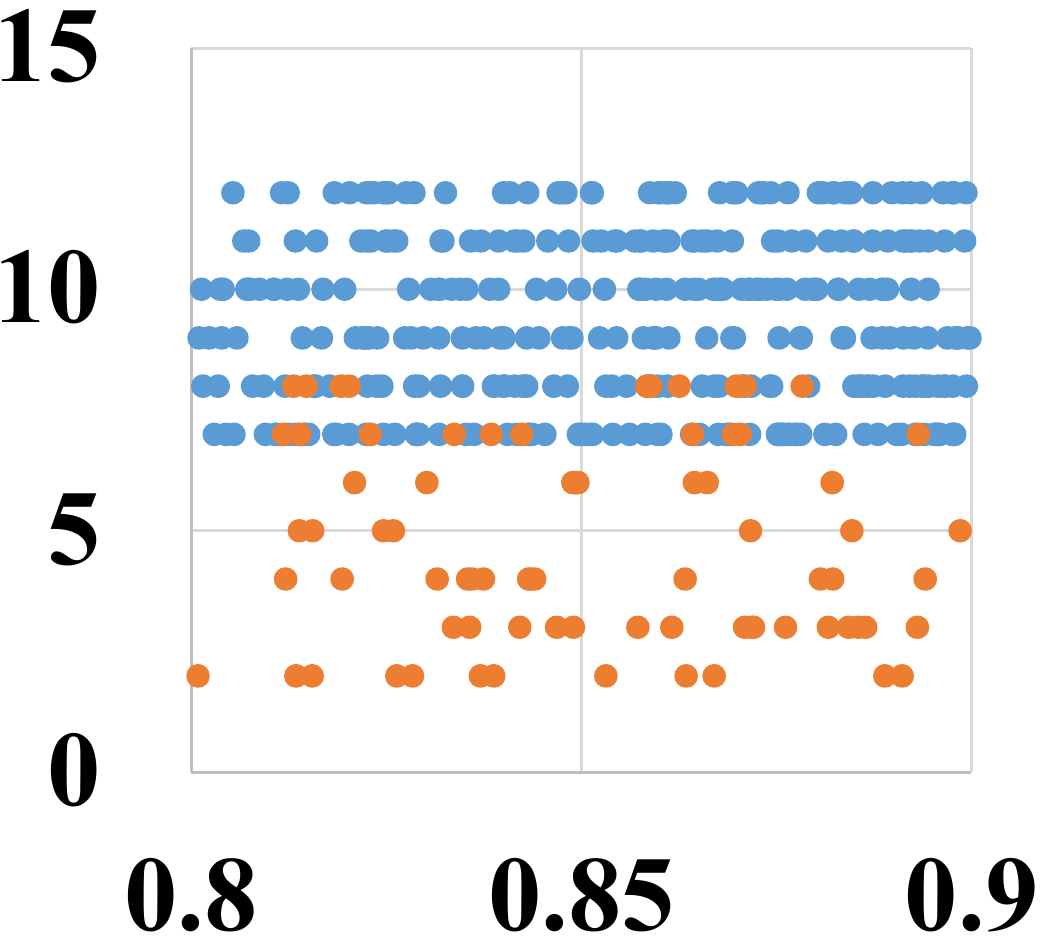}}
    \subfigure[\footnotesize[0.9,1.0)]{\includegraphics[width=0.19\linewidth,height=0.16\linewidth]{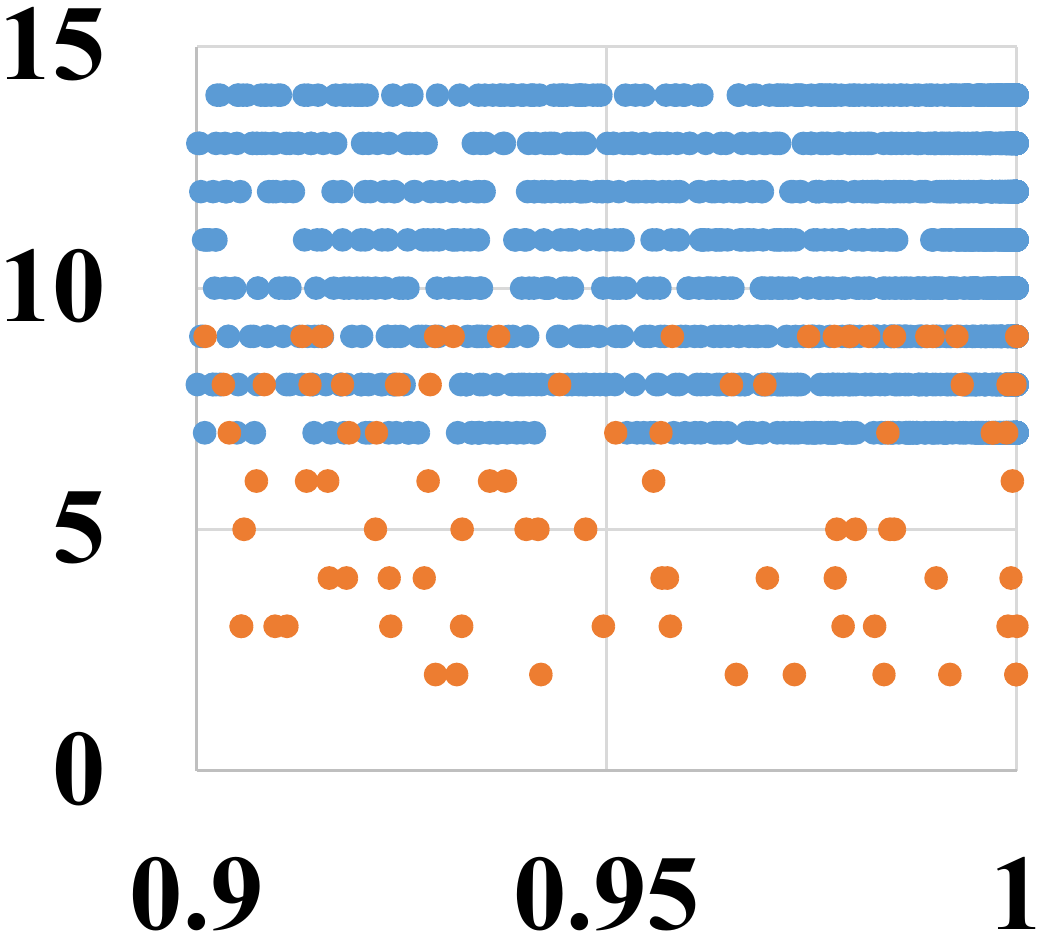}}
    \caption{An example of movement cost for test inputs with different probability levels,
    where ``\textcolor[RGB]{90,154,213}{$\bullet$}'' and ``\textcolor[RGB]{237,123,45}{$\bullet$}'' represent correctly and incorrectly predicted test inputs, respectively.
    Subfigure (a) represents movement costs required for test inputs whose top-1 prediction probability belongs to (0, 0.1) to reach a position with a probability higher than 0.1, 
    where the \textit{x}-axis represents the original top-1 prediction probability and 
    the \textit{y}-axis represents the movement cost.
    We repeat 5 times and compute the average movement cost of each test input.
    }
    \label{fig:movement_cost}
\end{figure}

One of the main contributions of DNNs is automatic feature extraction~\cite{LeCun2015DeepLearning}, 
which maps test inputs from the data space to the feature space.
Based on the mapping ability, 
Zheng \textit{et al}.~\cite{zheng2021grip} improved the robustness of DNNs by pushing the test input to the target position (i.e., class center) based on inverse perturbation.
Further, 
we find that the inverse perturbation measures the movement cost of test inputs in feature space. 
Thus, 
we explore the variation in the movement cost of different test inputs and give an example as shown in Figure~\ref{fig:movement_cost}.
We compute the movement cost (i.e., inverse perturbation based on infinite norm) of 5,000 test inputs from ImageNet~\cite{ILSVRC15ImageNet} on a pre-trained VGG model~\cite{Simonyan2015VGGNet}. 
We first divide test inputs into 10 groups according to the original prediction probability, 
and then calculate the movement cost of reaching a position with a higher probability.
As shown in Figure~\ref{fig:movement_cost}, 
we find that there was a significant difference (p-value=2.38E-07 based on T-test) in the movement cost of correctly and incorrectly predicted test inputs, 
which can be considered for prioritization.
However, 
the inverse perturbation~\cite{zheng2021grip} is obtained through iterative training, 
which is still empirical. 
To satisfy the certifiability requirement, 
we further derive a formal guarantee of the inverse perturbation with the Lipschitz continuity assumption~\cite{paulavivcius2006analysis}.


According to the utility analysis and certifiability consideration,
we design a certifiable prioritization technique, \emph{CertPri},
which reduces the problem of measuring misbehavior probability to the problem of measuring the movement difficulty in feature space,
i.e., the movement cost of the test inputs being close to or far from the class centers.
Then, 
we compute the certifiable inverse perturbation based on the generalized extreme value theory (GEVT)~\cite{Gnedenko1943SurLa}.
Based on the formal robustness guarantee,
CertPri is valid for identifying malicious bug-revealing inputs, 
as well as clean bug-revealing inputs,
without degrading efficiency.


To compute the inverse perturbation, 
the model gradient is used~\cite{Wei2018CLEVER}.
Since DNNs are an end-to-end learning paradigm~\cite{Tampuu2022EndToEnd}, 
the gradient can be directly computed when data and models are available in a white-box scenario.
In a black-box scenario, 
the approximate gradient can be computed by gradient estimation~\cite{Chen2017ZOO}. 
Furthermore, 
to evaluate CertPri's generalizability,
we conduct extensive experiments on
various tasks, data forms, data types, model structures, training and prioritization scenarios.


The main contributions are as follows.
\begin{itemize}
  \item Through inverse perturbation analysis and measurement,
        we first implement a formal robustness guarantee for the movement cost,
        which provides a new perspective for measuring DNNs' misbehavior probability.
  \item Based on the formal guarantee of movement costs,
        we propose an effective and efficient technique,
        CertPri,
        which leverages certifiability to facilitate the prioritization effectiveness.
  \item To evaluate CertPri's generalizability,
        we conduct extensive experiments on 
        various tasks, data, models and scenarios,
        the results of which show the superiority of CertPri compared with previous works.
  \item We publish CertPri as a self-contained open-source toolkit online 
        for facilitating DNNs' prioritization research.
\end{itemize}


\section{Background\label{Background}}
In this section, 
we first introduce the basic knowledge of DNNs,
and then give the definitions of inverse perturbation.

\subsection{Deep Neural Networks}
DNN consists of several layers,
each of which contains a large number of neurons\cite{LeCun2015DeepLearning}.
Generally,
the basic tasks of DNNs include classification and regression.
These tasks are accomplished by building models with various basic structures,
such as 
fully connected network (FCN)~\cite{Zheng2022NeuronFair},
convolutional neural network (CNN)~\cite{Simonyan2015VGGNet}, 
long short-term memory (LSTM)~\cite{Du2021CertRNN}, and
GCN~\cite{Zhang2020GraphsSurvey}.
The classification and regression models are as follows.

The classification model predicts which class the test input belongs to.
Suppose we have a $K$-class classifier $f^{C}:\mathbb{R}^{d}\to\mathbb{R}^{K}$.
Given a test input $\bm{x}_{0}\in\mathbb{R}^{d}$,
the classifier will output a vector of $K$ values normalized by softmax function~\cite{John1990Softmax},
e.g.,
$\{f_{i}^{C}(\bm{x}_{0})|_{1\le i\le K}\}$,
each of which represents the probability that $\bm{x}_{0}$ belongs to the $i$-th class,
where $\{d,K\}\in\mathbb{Z}^{+}$ and $K\ge$2.
$c(\bm{x}_{0})=\arg\max_{1\le i\le K} f_{i}^{C}(\bm{x}_{0})$ represents the predicted label of $\bm{x}_{0}$ and
$f_{i}^{C} (\bm{x}_{0}) \in (0,1)$.

The regression model describes a mapping between the test input and the output. 
Suppose we have a regression model $f^{R}:\mathbb{R}^{d_{1}}\to\mathbb{R}^{d_{2}}$.
Given a test input $\bm{x}_{0}\in\mathbb{R}^{d_{1}}$, 
the regression model will output a vector with $d_{2}$ elements activated by linear or ReLU functions~\cite{Konstantin2019ReLU},
e.g.,
$\{f_{i}^{R}(\bm{x}_{0})|_{1\le i\le d_{2}}\}$,
each of which represents a fit to the ground-truth,
where $\{d_{1},d_{2}\}\in\mathbb{Z}^{+}$.
$r(\bm{x}_{0})=f^{R}(\bm{x}_{0})$ represents the fitted prediction output of $\bm{x}_{0}$ and 
$f_{i}^{R}(\bm{x}_{0})\in [\min_{r}, \max_{r}]$.

\subsection{Definitions of Inverse Perturbation}
We give definitions of 
inversely perturbed test input, 
minimum inverse perturbation, and 
lower bound.

\textbf{Definition 1}
(inversely perturbed test input).
Given $\bm{x}_{0}$,
we say $\bm{x}_{0}^{*}$ is an inversely perturbed test input of $\bm{x}_{0}$ with inverse perturbation $\bm{\mu}$ and
$l_{p}$-norm $\Delta_{p}$ if $\bm{x}_{0}^{*} = \bm{x}_{0} + \bm{\mu}$ is moved to the target position and
$\Delta_{p} = ||\bm{\mu}||_{p}$.
An inversely perturbed test input for a classifier is $\bm{x}_{0}^{*}\in\mathbb{R}^{d}$ that moves towards the class center of $c(\bm{x}_{0})$,
where the class center is defined as $f_{center}^{C}(\bm{x}_{0}) = \min \left\{f_{c}^{C}(\bm{x}_{0})\times[1+\log(1+f_{c}^{C}(\bm{x}_{0}) )] , 1\right\}$.
For a regression model,
$\bm{x}_{0}^{*}\in\mathbb{R}^{d_{1}}$ moves from $r(\bm{x}_{0})$ to regression center,
which is defined as 
$f^{R}_{i,+}(\bm{x}_{0}) = {\rm clip}_{\min_{r}}^{\max_{r}} \left( f_{i}^{R}(\bm{x}_{0}) + |f_{i}^{R}(\bm{x}_{0})| \times \log \left[ 1 + \tanh( f_{i}^{R}(\bm{x}_{0}) ) \right]  \right)$.

\textbf{Definition 2}
(minimum inverse perturbation $\Delta_{p}^{\min}$ and its lower bound $\gamma_{L}$).
Given a test input $\bm{x}_{0}$,
the minimum $l_{p}$ inverse perturbation of $\bm{x}_{0}$, 
denoted as $\Delta_{p}^{\min}$, 
is defined as the smallest $\Delta_{p}$ over all inversely perturbed test inputs of $\bm{x}_{0}$.
Suppose $\Delta_{p}^{\min}$ is the minimum inverse perturbation of $\bm{x}_{0}$. 
A lower bound of $\Delta_{p}^{\min}$, 
denoted by $\gamma_{L}$ where $\gamma_{L} \le \Delta_{p}^{\min}$, 
is defined such that any inversely perturbed test inputs of $\bm{x}_{0}$ with $||\bm{\mu}||_{p}\le \gamma_{L}$ will never reach the target position.

The lower bound of inverse perturbation measures the minimum movement cost for a test input to reach the target position (i.e., class center or regression center).
$\gamma_{L}$ guarantees that the inversely perturbed test input will never move to the target position for inverse perturbation with $||\bm{\mu}||_{p} \le \gamma_{L}$,
certifying the movement cost of the test input. 
All the notations are summarized in Table~\ref{tab:Notations}.

\begin{table}[t]
    \centering
    \caption{Definitions of Notations.}
    \label{tab:Notations}
    \resizebox{\linewidth}{!}{
    \begin{tabular}{lll}
    \toprule \hline
                                      & \textbf{Notations}                                                                    & \textbf{Definitions}                                                                     \\ \hline
\multirow{10}{*}{\rotatebox{90}{\textbf{DNNs}}}                & $d$, $K$                                                                     & dimensionality of test inputs, number of output classes                         \\
                                      & $f^{C}: \mathbb{R}^{d}\to\mathbb{R}^{K}$                                     & classification model                                                            \\
                                      & $f^{C}_{i}(\bm{x}_{0})$                                                      & the $i$-th dimension output of the classification model                         \\
                                      & $c(\bm{x}_{0})$                                          & the predicted label of $\bm{x}_{0}$                                             \\
                                      & $d_{1}$, $d_{2}$                                                             & dimensionality of test inputs and outputs                                       \\
                                      & $f^{R}: \mathbb{R}^{d_{1}}\to\mathbb{R}^{d{2}}$                              & regression model                                                                \\
                                      & $f^{R}_{i}(\bm{x}_{0})$                                                      & the $i$-th dimension output of the regression model                             \\
                                      & $r(\bm{x}_{0})$                                          & the fitted prediction output of $\bm{x}_{0}$                                    \\
                                      & $\max_{r}$, $\min_{r}$                                                       & maximum and minimum values of regression output domain 
                                                        \\
                                      & $\bm{x}_{0}\in \mathbb{R}^{d}$, $\bm{x}_{0}\in \mathbb{R}^{d_{1}}$           & original test input of classification and regression              \\ \cline{2-3}
\multirow{7}{*}{\rotatebox{90}{\textbf{Inverse Perturbation}}} & $\bm{x}^{*}_{0}\in \mathbb{R}^{d}$, $\bm{x}^{*}_{0}\in   \mathbb{R}^{d_{1}}$ & inversely perturbed test input of classification and regression \\
                                      & $\bm{\mu} \in \mathbb{R}^{d}$, $\bm{\mu} \in \mathbb{R}^{d_{1}}$             & inverse perturbation of classification and regression             \\
                                      & $||\bm{\mu}||_{p}$  &  $l_{p}$-norm of inverse perturbation, $p\ge1$
                                              \\
                                      & $f^{C}_{center}(\bm{x}_{0})$, $f^{R}_{i,+}(\bm{x}_{0})$                      & class center and regression center                                              \\
                                      & $\min\{ \cdot, \cdot \}$                                                     & returns the minimum element in the set                                          \\
                                      & ${\rm clip}_{\min_{r}}^{\max_{r}} (\cdot)$                                   & limit the input value to the range of ($\min_{r}$, $\max_{r}$)                  \\
                                      & $\Delta_{p}^{min}$, $\gamma_{L}$                                             & minimum inverse perturbation and its lower bound                                  \\
                                      & $L_{q}^{c}$, $L_{q}^{r}$                                                     & Lipschitz constant of classification and regression, $q\ge1$   \\                  \hline \bottomrule
    \end{tabular}
}
\end{table}

\section{CertPri Methodology\label{Method}}
In this section, we present a technical description of CertPri. 
First, we discuss the prioritization feasibility based on a movement cost view.
Then, we introduce CertPri,
which provides 
a formal robustness guarantee based on the Lipschitz continuity assumption~\cite{paulavivcius2006analysis} and 
estimates the movement cost based on GEVT~\cite{Wei2018CLEVER}.
Finally, we prioritize inputs via movement costs.

\subsection{A Movement View in Feature Space\label{MovementView}}

A well-trained DNN implements feature extraction through multiple hidden layers, 
each of which filters redundant features and amplifies key features during forward propagation~\cite{chen2021act}.
If we regard the feature mapping of hidden layers as data movement in feature space, 
almost all test inputs are pushed towards the target position in forward propagation, 
while bug-revealing inputs fail to reach the target position at the end of forward propagation.

\begin{figure}[t]
    \centering
    \subfigure[The movement process of test inputs in forward propagation,
    where almost all test inputs move towards the corresponding class centers.]{
        \includegraphics[width=0.3\linewidth]{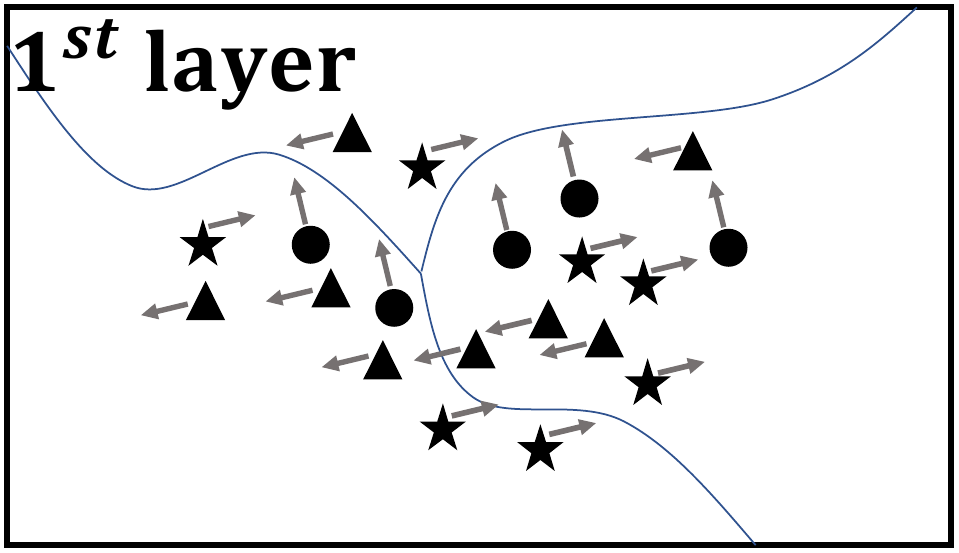}
        \includegraphics[width=0.3\linewidth]{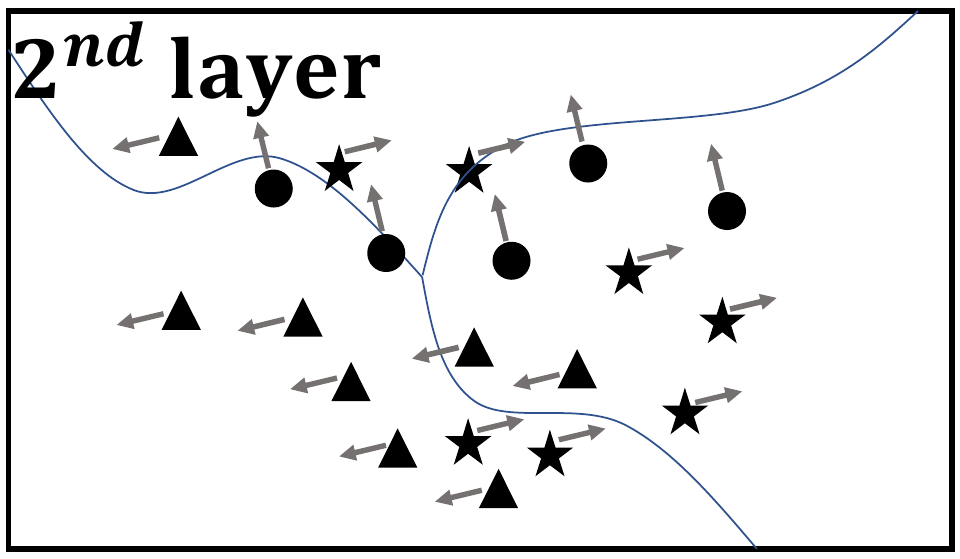}
        \includegraphics[width=0.3\linewidth]{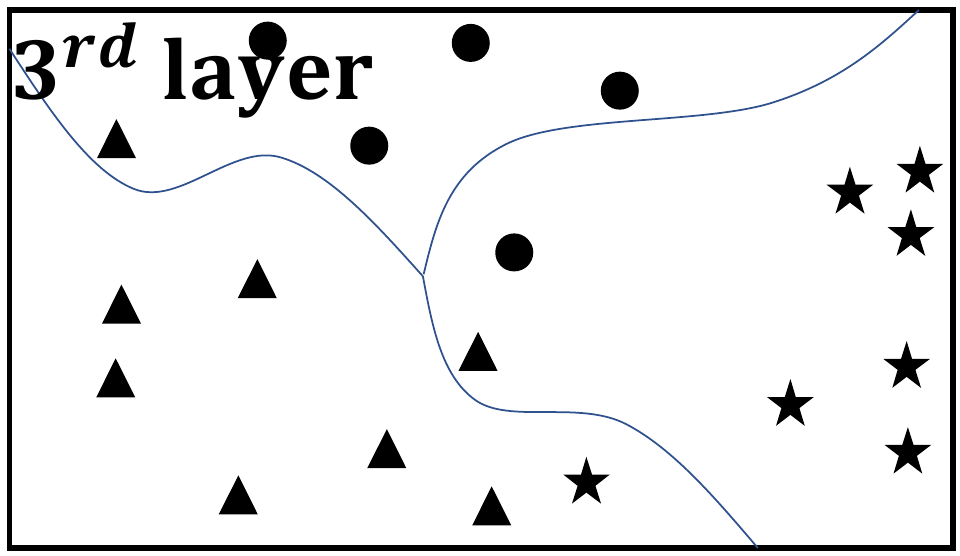} 
        } \\
    \subfigure[The movement cost of test inputs in back propagation,
    where $\bm{x}_{0}$ is a misclassified test input,
    ``probability contour'' represents that test inputs on the same contour line have the same probability of belonging to the class,
    ``gradient direction'' is obtained by back propagation~\cite{Amari1993Backpropagation}.
    ]{
        \includegraphics[width=0.91\linewidth]{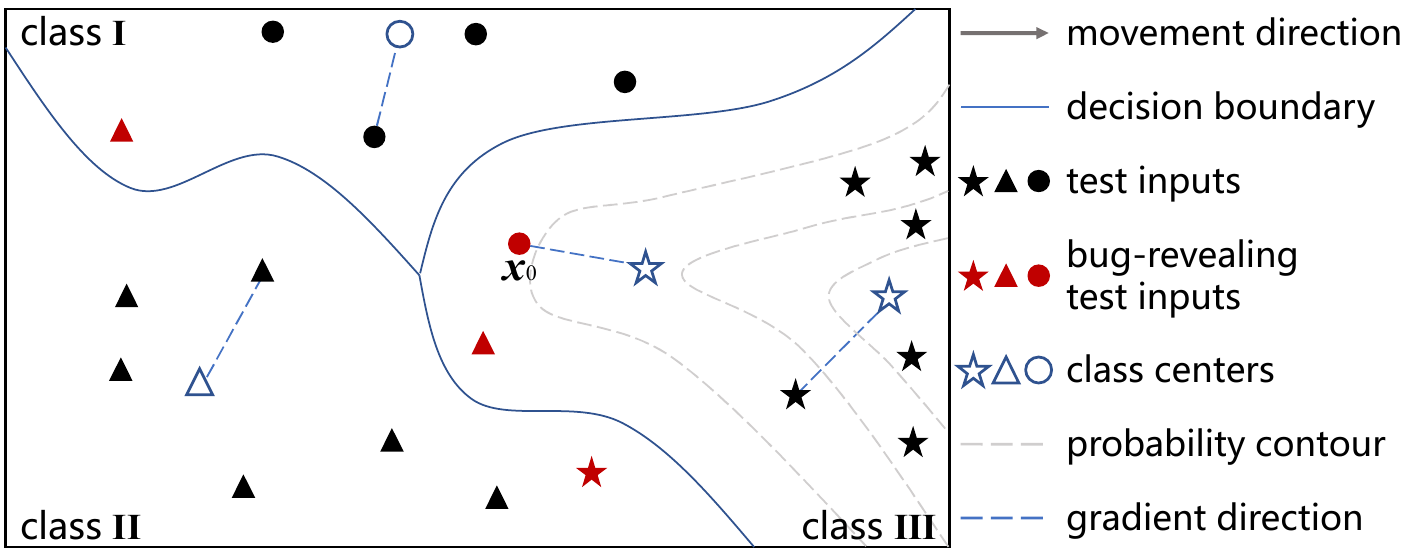}
        }
    \caption{An example of t-SNE~\cite{Van2009tSNE} based distribution visualization of test inputs in feature space.
    The classifier is a FCN with three hidden layers,
    the test inputs are handwritten digits of ``0''(I), ``1''(II) and ``2''(III) from MNIST dataset~\cite{LeCun1989MNIST}.
    }
    \label{fig:movement_view}
\end{figure}

First, we investigate the movement process of test inputs in forward propagation.
Taking a classifier based on a FCN with three hidden layers on MNIST~\cite{LeCun1989MNIST} dataset as an example, 
the t-SNE~\cite{Van2009tSNE} based distribution of test inputs in feature space is visualized in Figure~\ref{fig:movement_view}.
During the forward propagation as shown in Figure~\ref{fig:movement_view}~(a), 
we observe that most test inputs directional approach the correct class, 
which realizes the right prediction.
However,
several test inputs move without direction leading to the DNN's misbehavior (i.e., misclassification),
which are the bug-revealing inputs that need to be identified.
It is intuitively based on the feature purity theory of test inputs~\cite{Feng2020DeepGini}.
For a bug-revealing test input,
it not only contains features of multiple classes, 
but the highest feature purity is close to the second highest one, 
and even the feature purity of each class is close.
Thus,
the movement of this test input in the forward propagation is directionless.

Then,
we further analyze the movement cost of correctly and incorrectly predicted test inputs based on the inverse perturbation~\cite{zheng2021grip}.
As shown in Figure~\ref{fig:movement_view}~(b), 
the ground truth of $\bm{x}_{0}$ is \textbf{II} but is misclassified as \textbf{III}.
As $\bm{x}_{0}$ proceeds along the gradient direction, 
its class center of \textbf{III} can be reached with low movement cost. 
However, correctly predicted test inputs require high movement cost to reach their class centers. 
It is consistent with the interpretation based on feature purity~\cite{Feng2020DeepGini}. 
The incorrectly predicted test inputs improve the feature purity of the corresponding class after adding inverse perturbation, 
which turns random movement into directional movement in forward propagation.
Therefore, only a low movement cost is required to reach the class center. 
The correctly predicted test inputs contain high feature purity,
which keeps them stable in feature space.
Thus, 
further movement requires a high cost.
Note that the class center is given by \textbf{Definition 1},
which differs for each test input~\cite{zheng2021grip}.

Based on the above analysis,
we reduce the problem of measuring misbehavior probability of prioritization to the problem of measuring the movement cost in feature space.
Thus,
we prioritize test inputs by comparing their lower bounds of minimum movement cost to the target position.
The test input never reaches the target position when the movement cost is less than the lower bound $\gamma_{L}$.
However,
$\gamma_{L}$ is not easy to find.
Below we show how to derive a formal inverse perturbation guarantee of a test input with the Lipschitz continuity assumption~\cite{paulavivcius2006analysis}. 

\subsection{Formal Guarantees for Movement Cost\label{FormalGuarantees}}

We first give the lemma about the Lipschitz continuity. 
According to the lemma, 
we then provide a formal guarantee for the lower bound of the inverse perturbation.
Specifically, 
our analysis obtains a lower bound of $l_{p}$-norm minimum inverse perturbation
$\gamma_{L}=\min\frac{f_{center}^{C}(\bm{x}_{0}) - f_{c}^{C}(\bm{x}_{0})}{L_{q}^{c}}$ for a classifier and 
$\gamma_{L}\!=\!\min\frac{ \sum_{i}| f_{i,+}^{R}(\bm{x}_{0}) - f_{i}^{R}(\bm{x}_{0})| } {d_{2} \times L_{q}^{r}}$ for a regression model.

\textbf{Lemma 1}
(Norms and corresponding Lipschitz constants~\cite{paulavivcius2006analysis}).
\textit{
Let $D\subset\mathbb{R}^{d}$ be a convex bounded closed set and let $h(\bm{x}):S\to\mathbb{R}$ be a continuously differentiable function on an open set containing $D$.
For a Lipschitz function $h(\bm{x})$ with Lipschitz constant $L_{q}$,
the inequality $|h(\bm{a})-h(\bm{b})|\le L_{q}||\bm{a}-\bm{b}||_{p}$ holds for any $\{\bm{a},\bm{b}\}\in D$,
where $L_{q}=\max\{ ||\nabla_{\bm{x}} h(\bm{x})||_{q} : \bm{x}\in D \}$,
$\nabla_{\bm{x}} h(\bm{x}) = ( \frac{\partial h}{\partial \bm{x}_{1}}, ..., \frac{\partial h}{\partial \bm{x}_{d}}  )$
is the gradient of $h(\bm{x})$,
$\frac{1}{p} + \frac{1}{q} = 1$ and
$1\le\{p,q\}\le\infty$.
}

\textbf{Theorem 1}
(Formal guarantee on lower bound $\gamma_{L}$ of inverse perturbation for classification model).
\textit{Let $\bm{x}_{0}\in\mathbb{R}^{d}$ and 
$f^{C}:\mathbb{R}^{d}\to\mathbb{R}^{K}$ be a $K$-class classifier with continuously differentiable components.
For all $\bm{\mu}\in\mathbb{R}^{d}$ with
$||\bm{\mu}||_{p}\le\min\frac{f_{center}^{C}(\bm{x}_{0}) - f_{c}^{C}(\bm{x}_{0})}{L_{q}^{c}}$,
$f_{c}^{C}( \bm{x}_{0} + \bm{\mu} ) = f_{center}^{C}(\bm{x}_{0})$ holds with 
$\frac{1}{p} + \frac{1}{q} = 1$,
$1\le\{p,q\}\le\infty$ and
$L_{q}^{c}$ is the Lipschitz constant for the function $f_{center}^{C}(\bm{x})-f_{c}^{C}(\bm{x})$ in $l_{p}$-norm.
In other word,
$\gamma_{L}=\min\frac{f_{center}^{C}(\bm{x}_{0}) - f_{c}^{C}(\bm{x}_{0})}{L_{q}^{c}}$ is a lower bound of minimum inverse perturbation for moving to the class center.
}

\textit{\textbf{Proof}}.
According to Lemma 1,
the assumption that the function $h(\bm{x}) := f_{center}^{C}(\bm{x}) - f_{c}^{C}(\bm{x})$ is Lipschitz continuous with Lipschitz constant $L_{q}^{c}$ gives:
\begin{equation}
    |h(\bm{a})-h(\bm{b})|\le L_{q}^{c}||\bm{a}-\bm{b}||_{p}.
    \label{equ:proof_Lip_a}
\end{equation}

Let $\bm{a}=\bm{x}_{0}+\bm{\mu}$ and $\bm{b}=\bm{x}_{0}$,
we get:
\begin{equation}
    | h(\bm{x}_{0}+\bm{\mu}) - h(\bm{x}_{0}) | \le L_{q}^{c} || \bm{\mu} ||_{p},
\end{equation}
which can be rearranged by removing the absolute symbol: 
\begin{equation}
    \begin{array}{ll} 
         &\!\! -L_{q}^{c} ||\bm{\mu}||_{p} \le h(\bm{x}_{0}+\bm{\mu}) - h(\bm{x}_{0}) \le L_{q}^{c} ||\bm{\mu}||_{p}, \\
        \!\!\! \Rightarrow \!\!&\!\! h(\bm{x}_{0}) - L_{q}^{c} ||\bm{\mu}||_{p} \le h(\bm{x}_{0}+\bm{\mu}) \le h(\bm{x}_{0}) + L_{q}^{c} ||\bm{\mu}||_{p} .
    \end{array}
    \label{equ:rearranged}
\end{equation}

When $h(\bm{x}_{0}+\bm{\mu}) \!\!=\!\! 0$,
the inversely perturbed test input is moved to the class center.
As represented by Equation~(\ref{equ:rearranged}),
$h(\bm{x}_{0}) - L_{q}^{c} ||\bm{\mu}||_{p}$ is the lower bound of $h(\bm{x}_{0}+\bm{\mu})$.
If $h(\bm{x}_{0}) - L_{q}^{c} ||\bm{\mu}||_{p} \ge 0 $ for sufficiently small inverse perturbation $||\bm{\mu}||_{p}$,
the inversely perturbed input cannot reach the class center,
i.e.,
\begin{equation}
    \begin{array}{ll}
         \!\!\! h(\bm{x}_{0}) - L_{q}^{c} ||\bm{\mu}||_{p} \ge 0
        &\!\! \Rightarrow  ||\bm{\mu}||_{p} \le \frac{h(\bm{x}_{0})}{L_{q}^{c}} \\
        &\!\!\Rightarrow  ||\bm{\mu}||_{p} \le \frac{ f_{center}^{C}(\bm{x}_{0}) - f_{c}^{C}(\bm{x}_{0}) }{L_{q}^{c}}.
    \end{array}
\end{equation}

To realize $f_{c}^{C}( \bm{x}_{0} + \bm{\mu} ) = f_{center}^{C}(\bm{x}_{0})$
we take the minimum of the bound on $||\bm{\mu}||_{p}$,
i.e.,
the test input will never move to the class center when $||\bm{\mu}||_{p} \le \min \frac{f_{center}^{C}(\bm{x}_{0}) - f_{c}^{C}(\bm{x}_{0})}{L_{q}^{c}}$.

\textbf{Theorem 2}
(Formal guarantee on lower bound $\gamma_{L}$ of inverse perturbation for regression model).
\textit{Let $\bm{x}_{0}\in\mathbb{R}^{d_{1}}$ and 
$f^{R}:\mathbb{R}^{d_{1}}\to\mathbb{R}^{d_{2}}$ be a regression model with continuously differentiable components.
For all $\bm{\mu}\in\mathbb{R}^{d_{1}}$ with
$||\bm{\mu}||_{p}\le \min \frac{ \sum_{i} | f_{i,+}^{R}(\bm{x}_{0}) - f_{i}^{R}(\bm{x}_{0})| } {d_{2} \times L_{q}^{r}} $,
$\frac{1}{d_{2}} \sum | r( \bm{x}_{0} + \bm{\mu} ) - r(\bm{x}_{0})| \le \delta $ holds with 
$\frac{1}{p} + \frac{1}{q} = 1$,
$1\le\{p,q\}\le\infty$ and
$L_{q}^{r}$ is the Lipschitz constant for the function 
$\frac{ \sum_{i}| f_{i,+}^{R}(\bm{x}) - f_{i}^{R}(\bm{x})| } {d_{2}} $ in $l_{p}$-norm.
In other word,
$\gamma_{L}=\min\frac{ \sum_{i}| f_{i,+}^{R}(\bm{x}_{0}) - f_{i}^{R}(\bm{x}_{0})| } {d_{2} \times L_{q}^{r}}$ is a lower bound of minimum inverse perturbation.
}
The complete proof is deferred to \url{https://anonymous.4open.science/r/CertPri/SupplementaryMaterials.pdf}.

\begin{figure}[t]
    \centering
    \subfigure[PDF]{
        \includegraphics[width=0.45\linewidth]{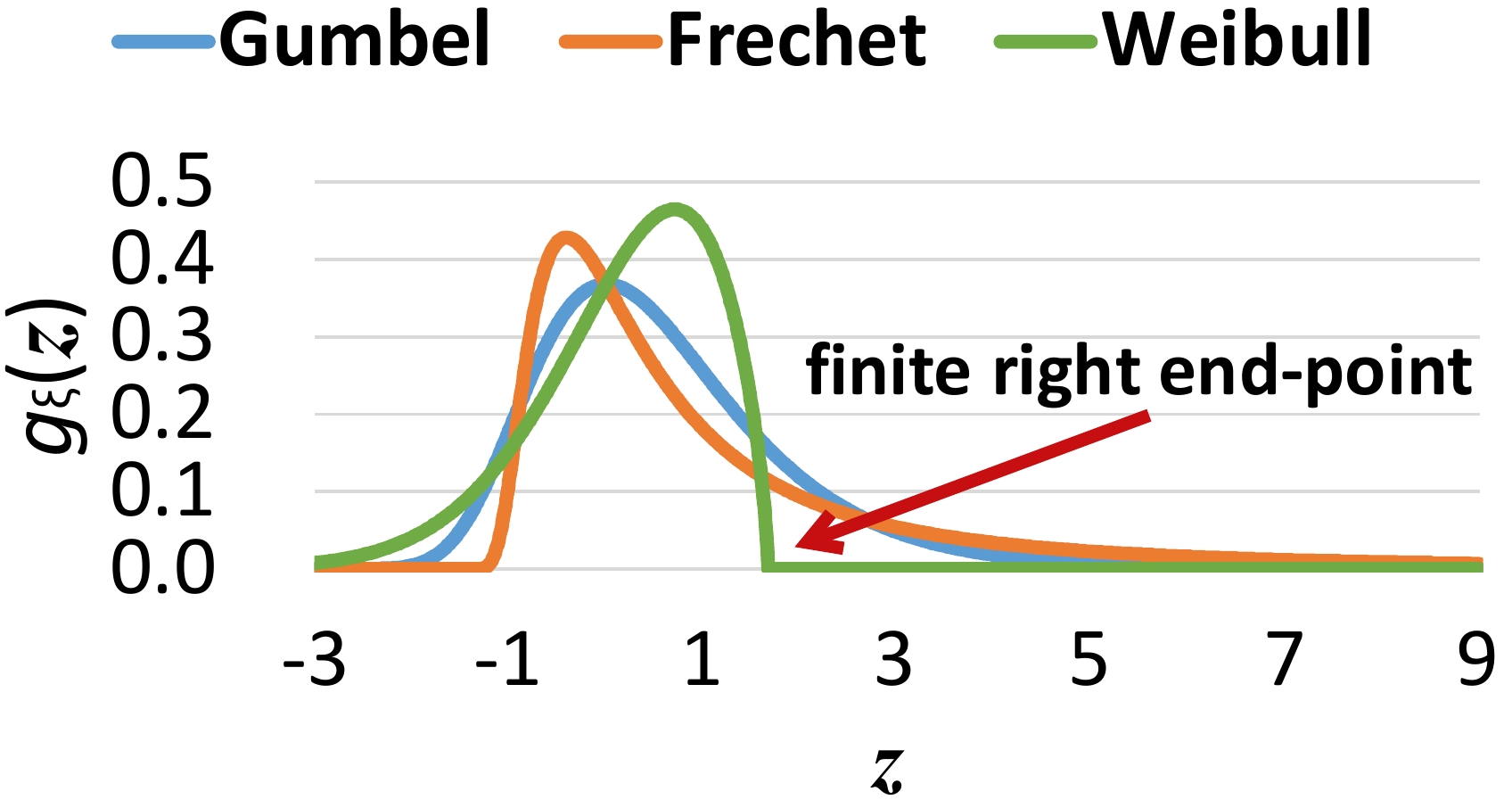}
        } 
    \subfigure[CDF]{
        \includegraphics[width=0.43\linewidth]{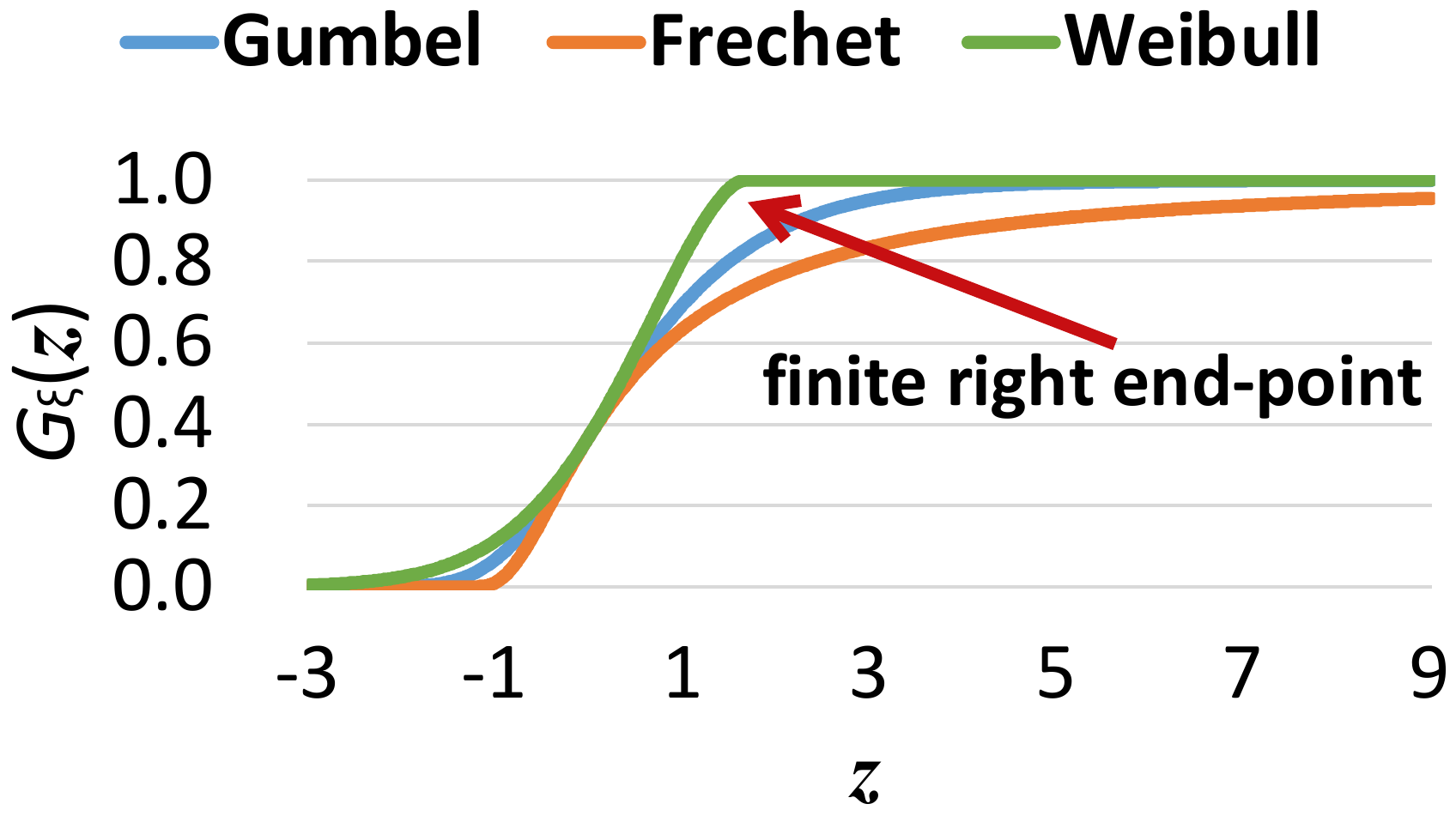}
        }
    \caption{Curves of PDF and CDF for different distributions,
    where $\xi=0$ for Gumbel, 
    $\xi=0.6$ for Fr$\rm{\acute{e}}$chet and 
    $\xi=-0.6$ for Weibull,
    $u=0$, $\sigma=1$.
    }
    \label{fig:PDF_CDF}
\end{figure}

\subsection{Movement Cost Estimation via GEVT} 
The formal guarantees give that 
$\gamma_{L}$ is related to $h(\bm{x}_{0})$ and its Lipschitz constant $L_{q}$,
where $L_{q}=\max\{ ||\nabla_{\bm{x}} h(\bm{x})||_{q} : \bm{x}\in B_{p}(\bm{x}_{0},\mathcal{R}) \}$,
$B_{p}(\bm{x}_{0},\mathcal{R}):=\{\bm{x}|_{||\bm{x}-\bm{x}{0}||_{p} \le \mathcal{R}} \}$ is a hyper-ball with center $\bm{x}_{0}$ and radius $\mathcal{R}$.
The value of $h(\bm{x}_{0})$ is easily accessible at the model output.
Thus,
we show how to get $\max ||\nabla_{\bm{x}} h(\bm{x})||_{q}$.

For a random variable sequence $\{ \bm{x}_{0}^{(j)} \}$ sampled from $B_{p}(\bm{x}_{0},\mathcal{R})$,
its corresponding gradient norm can be regarded as a new random variable sequence $\{ ||\nabla_{\bm{x}} h(\bm{x}_{0}^{(j)})||_{q} \}$ characterized by a cumulative distribution function (CDF)~\cite{Wei2018CLEVER}.
Therefore,
we estimate $\max ||\nabla_{\bm{x}} h(\bm{x}_{0}^{(j)})||_{q}$ with a small number of samples based on GEVT,
which ensures that the maximum value of a random variable sequence can only follow one of the three extreme value distributions~\cite{Gnedenko1943SurLa}.
The CDF of $||\nabla_{\bm{x}} h(\bm{x}_{0}^{(j)})||_{q}$ is as follows:
\begin{equation}
    G_{\xi}(\bm{z}) = 
            \exp\left( -( 1+\xi \bm{z} )^{-\frac{1}{\xi}} \right), 
\end{equation}
where $1\!+\!\xi\bm{z}>0$, 
$\bm{z}=\frac{ || \nabla_{\bm{x}} h(\bm{x}_{0}^{(j)})||_{q} - u}{\sigma}$,
$\xi\in \mathbb{R}$ is a extreme value index,
$u\in \mathbb{R}$ and $\sigma \in \mathbb{R}^{+}$ are the expectation and variance.
The parameters $u$, $\sigma$ and $\xi$ determine the location, scale and shape of $G_{\xi}(\bm{z})$, respectively.
$G_{\xi}(\bm{z})$ belongs to either the 
Gumbel ($\xi=0$),
Fr$\rm{\acute{e}}$chet ($\xi>0$) or
Weibull ($\xi<0$) distributions.

We plot curves of probability density function (PDF) and its CDF for different extreme value distributions in Figure~\ref{fig:PDF_CDF},
where PDF (denoted as $g_{\xi}(\bm{z})$) is the differential of CDF. 
The Weibull distribution shows an interesting property in Figure~\ref{fig:PDF_CDF},
i.e., a finite right end-point (denoted as $-\frac{1}{\xi}$),
which limits the upper bound of the distribution.
Therefore, 
we adopt the Weibull to describe the gradient norm distribution, 
where the right end-point is the estimation of $\max ||\nabla_{\bm{x}} h(\bm{x})||_{q}$.

{
\renewcommand\arraystretch{0.92}
\begin{table}[t]
  \centering
  \begin{tabular}{lp{0.86\columnwidth}}
  \toprule
    \multicolumn{2}{l} {\textbf{Algorithm 1}:
    CertPri.
    } \\ \hline
      &
    \textbf{Input}: Test inputs $\{\bm{x}_{i}\}$,
                    $i\in\{0,1,...,N\!-\!1\}$,
                    a classification model $f^{C}$, 
                    the sampling radius $\mathcal{R}$, 
                    batch size $N_{b}$, 
                    number of random samples per batch $N_{{rsb}}$, 
                    norm value $p$.
    \\
      &
    \textbf{Output}: The prioritization index $\Omega$.
    \\ \cline{2-2}
    1 &
    $h(\bm{x}):=f^{C}_{center}(\bm{x}) - f^{C}_{c}(\bm{x})$, 
    $q=\frac{p}{p-1}$, 
    $\Gamma=\{\emptyset\}$.
    \\
    2   & \textbf{For} $i=0:N\!-\!1$ \textbf{do}
    \\
    3   & \quad $S_{\nabla} = \{\emptyset\}$.
    \\
    4   & \quad \textbf{For} $j=0: N_{b}\!-\!1$ \textbf{do}
    \\
    5   & \quad\quad \textbf{For} $k=0: N_{rsb}\!-\!1$ \textbf{do}
    \\
    6   & \quad\quad\quad $g_{i,j,k}\!=\!||\nabla_{\bm{x}} h(\bm{x}_{i}^{(j,k)}) ||_{q}$.
    \\
    7   & \quad\quad \textbf{End For}
    \\
    8   & \quad\quad $g_{i,j}^{\max} = \max\{ g_{i,j,k} \}$ and $S_{\nabla} = S_{\nabla}\cup \{ g_{i,j}^{\max} \}$.
    \\
    9   & \quad \textbf{End For}
    \\
    10  & \quad Estimate $\hat{\xi}$ of Weibull distribution on $S_{\nabla}$.
    \\
    11  & \quad $\gamma_{L} = \frac{ h(\bm{x}_{i}) }{-1/\hat{\xi}}$ and $\Gamma = \Gamma \cup {\gamma_{L}}$.
    \\
    12  & \textbf{End For}
    \\
    13  & $\Omega$ $\gets$ return the index of $\Gamma$ in ascending order.
    \\
  \bottomrule
  \end{tabular}
\end{table}
}

\subsection{Prioritization through Movement Cost} 

Based on the movement cost perspective in Section~\ref{MovementView}, 
we can conclude that 
a test input with a small value of $\gamma_{L}$ should be prioritized at the front. 
Therefore,
we compute the $\gamma_{L}$ value of each test input,
and then prioritize these inputs according to their $\gamma_{L}$ values from small to large.
\textbf{Algorithm 1} shows the details of CertPri by taking the classification model as an example.
We first establish the gradient norm distribution of a test input by randomly sampling in the hyper-ball (the loop from lines 4 to 9).
Then,
we estimate the location of the maximum gradient norm based on the Weibull distribution and 
compute the lower bound of the movement cost based on \textbf{Theorem 1}
(lines 10 and 11).
Finally,
we repeat the above operations for each test input (the loop from lines 2 to 12) and 
prioritize them according to the ascending result of their $\gamma_{L}$ values
(line 13).
Note that for algorithmic illustration,
we only compute one $g_{i,j}^{\max}$ (line 8) for each iteration. 
To implement the best efficiency of GPU, 
we usually evaluate these values in batches,
and thus a batch of $g_{i,j}^{\max}$ can be returned.

Furthermore, 
we can easily generalize CertPri to a regression model through replacing the $h(\bm{x})$ function at line 1 with $h(\bm{x}) := \frac{ \sum_{i} | f_{i,+}^{R}(\bm{x}) - f_{i}^{R}(\bm{x})| } {d_{2}}$ based on \textbf{Theorem 2}.
Besides,
we can extend CertPri to a black-box scenario based on the gradient estimation~\cite{Chen2017ZOO}.

\section{Experimental Setting\label{Setting}}
In this section,
we introduce the experimental setup,
including the subjects we considered,
the baselines we compared with,
the measurements we used, and
the implementation details we conducted.

\subsection{Subjects}
We adopt 50 pairs of datasets and models as subjects,
as shown in Table~\ref{tab:subjects}. 
To sufficiently evaluate CertPri's effectiveness, efficiency, robustness, generalizability and guidance,
we carefully consider the diversity of subjects from six dimensions. 
To our best knowledge, 
this is the most large-scale and diverse study in the field.

(1) \textbf{Various tasks of deep models}.
We employ both classification models (ID: 1-32, 37-45, 47-49) and
regression models (ID: 33-36, 46, 50). 
The number of classes ranges from 2 to 1,000 across all the classification tasks.

(2) \textbf{Various data forms of test inputs}.
We consider six data forms of test inputs from 14 datasets, including
image, text, speech, signal, graph, and structured data.
Specifically,
we collect \textit{6 image datasets}, i.e.,
CIFAR10~\cite{krizhevsky2009learningCIFAR10}
(a 10-class ubiquitous object dataset with 32$\times$32 pixels),
ImageNet~\cite{ILSVRC15ImageNet} 
(a 1,000-class object recognition benchmark with 224$\times$224 pixels),
DrivingSA~\cite{Deng2020AutonomousDriving}
(a steer angle regression dataset for autonomous driving from Udacity with 128$\times$128 pixels),
Fashion-MNIST (FMNIST)~\cite{xiao2017FMNIST} 
(a 10-class dataset of Zalando's article images with 28$\times$28 pixels),
Ants\_Bees (a binary insect dataset containing ants and bees for transfer learning with 224$\times$224 pixels),
and Cats\_Dogs (a binary pet dataset containing cats and dogs for transfer learning with 224$\times$224 pixels),
\textit{2 text datasets}, i.e.,
IMDB~\cite{Maas2011LearningIMDB}
(a binary movie review sentiment dataset encoded as a list of word indexes)
and Reuters (a 46-class newswire dataset),
\textit{one speech dataset}, i.e.,
VCTK10
(a 10-class dataset of ten English speakers with various accents selected from VCTK Corpus with 601$\times$64 phonemes),
\textit{one signal dataset}, i.e.,
RML8PSK~\cite{2016Radio} 
(a signal regression dataset modulated by 8PSK from RadioML 2016.10a with 127$\times$2 points as input and 2 points as output),
\textit{one graph dataset}, i.e.,
Cora~\cite{Andrew2000AutomatingCora}
(a 7-class citation network of scientific publications with 5,429 links),
and \textit{3 structured datasets}, i.e.,
Adult~\cite{Kohavi1996ScalingAdult} 
(a binary income dataset with 13-dimensional attributes),
COMPAS 
(a binary crime prediction dataset with 400-dimensional attributes),
and Boston 
(a housing price regression dataset with 13-dimensional input and 1-dimensional output).

\begin{table}[ht]
\centering
\caption{Basic information of subjects.}
\label{tab:subjects}
\resizebox{\linewidth}{!}{\Huge
\begin{tabular}{l|lllrrrrr}
\toprule
\hline
\textbf{ID} & \textbf{Datasets} & \textbf{Models} & \textbf{Struc.} & \textbf{\#Inputs} & \textbf{Types} & \textbf{Forms} & \textbf{Tasks} & \textbf{Scenarios} \\ \hline
1           & CIFAR10           & ResNet50           & CNN             & 10,000            & original       & image          & C             & N+W           \\
2           & CIFAR10           & ResNet50           & CNN             & 10,000            & original       & image          & C             & {N+B }           \\
3           & CIFAR10           & ResNet50           & CNN             & 10,000            & +BIM           & image          & C             & N+W           \\
4           & CIFAR10           & ResNet50           & CNN             & 10,000            & +C\&W          & image          & C             & N+W           \\
5           & CIFAR10           & ResNet50           & CNN             & 10,000            & +FineFool      & image          & C             & N+W           \\
6           & CIFAR10           & ResNet50           & CNN             & 10,000            & +AdapS         & image          & C             & N+W           \\
7           & CIFAR10           & ResNet50           & CNN             & 10,000            & +AdapC         & image          & C             & N+W           \\
8           & CIFAR10           & ResNet50           & CNN             & 10,000            & +AdapM         & image          & C             & N+W           \\
9           & CIFAR10           & VGG16        & CNN             & 10,000            & original       & image          & C             & N+W           \\
10          & CIFAR10           & VGG16        & CNN             & 10,000            & original       & image          & C             & {N+B }           \\
11          & CIFAR10           & VGG16        & CNN             & 10,000            & +BIM           & image          & C             & N+W           \\
12          & CIFAR10           & VGG16        & CNN             & 10,000            & +C\&W          & image          & C             & N+W           \\
13          & CIFAR10           & VGG16        & CNN             & 10,000            & +FineFool      & image          & C             & N+W           \\
14          & CIFAR10           & VGG16        & CNN             & 10,000            & +AdapS         & image          & C             & N+W           \\
15          & CIFAR10           & VGG16        & CNN             & 10,000            & +AdapC         & image          & C             & N+W           \\
16          & CIFAR10           & VGG16        & CNN             & 10,000            & +AdapM         & image          & C             & N+W           \\
17          & ImageNet          & ResNet101           & CNN             & 5,000             & original       & image          & C             & N+W           \\
18          & ImageNet          & ResNet101           & CNN             & 5,000             & original       & image          & C             & {N+B }           \\
19          & ImageNet          & ResNet101           & CNN             & 5,000             & +BIM           & image          & C             & N+W           \\
20          & ImageNet          & ResNet101           & CNN             & 5,000             & +C\&W          & image          & C             & N+W           \\
21          & ImageNet          & ResNet101           & CNN             & 5,000             & +FineFool      & image          & C             & N+W           \\
22          & ImageNet          & ResNet101           & CNN             & 5,000             & +AdapS         & image          & C             & N+W           \\
23          & ImageNet          & ResNet101           & CNN             & 5,000             & +AdapC         & image          & C             & N+W           \\
24          & ImageNet          & ResNet101           & CNN             & 5,000             & +AdapM         & image          & C             & N+W           \\
25          & ImageNet          & VGG19       & CNN             & 5,000             & original       & image          & C             & N+W           \\
26          & ImageNet          & VGG19       & CNN             & 5,000             & original       & image          & C             & {N+B }           \\
27          & ImageNet          & VGG19       & CNN             & 5,000             & +BIM           & image          & C             & N+W           \\
28          & ImageNet          & VGG19       & CNN             & 5,000             & +C\&W          & image          & C             & N+W           \\
29          & ImageNet          & VGG19       & CNN             & 5,000             & +FineFool      & image          & C             & N+W           \\
30          & ImageNet          & VGG19       & CNN             & 5,000             & +AdapS         & image          & C             & N+W           \\
31          & ImageNet          & VGG19       & CNN             & 5,000             & +AdapC         & image          & C             & N+W           \\
32          & ImageNet          & VGG19       & CNN             & 5,000             & +AdapM         & image          & C             & N+W           \\
33          & DrivingSA         & VGG19-AD        & CNN             & 5,279             & original       & image          & R             & N+W           \\
34          & DrivingSA         & VGG19-AD        & CNN             & 5,279             & original       & image          & R             & {N+B }           \\
35          & DrivingSA         & VGG19-AD        & CNN             & 5,279             & patch          & image          & R             & N+W           \\
36          & DrivingSA         & VGG19-AD        & CNN             & 5,279             & saturation     & image          & R             & N+W           \\
37          & FMNIST            & AlexNet         & CNN             & 10,000            & original       & image          & C             & N+W           \\
38          & FMNIST\_P            & AlexNet-P       & CNN             & 10,000            & original       & image          & C             & P+W           \\
39          & Ants\_Bees        & VGG16-AB        & CNN             & 153               & original       & image          & C             & T+W           \\
40          & Cats\_Dogs        & VGG19-CD        & CNN             & 5,000             & original       & image          & C             & T+W           \\
41          & IMDB              & CNN-I           & CNN             & 10,000            & original       & text           & C             & N+W           \\
42          & IMDB              & LSTM-I          & LSTM            & 10,000            & original       & text           & C             & N+W           \\
43          & Reuters           & CNN-R           & CNN             & 2,246             & original       & text           & C             & N+W           \\
44          & Reuters           & LSTM-R          & LSTM            & 2,246             & original       & text           & C             & N+W           \\
45          & VCTK10            & LSTM-V          & LSTM            & 400               & original       & speech         & C             & N+W           \\
46          & RML8PSK           & LSTM-RML        & LSTM            & 312               & original       & signal         & R             & N+W           \\
47          & Cora              & GCN-C           & GCN             & 1,000             & original       & graph          & C             & N+W           \\
48          & Adult             & LFCN-A          & FCN             & 10,000            & original       & structured     & C             & N+W           \\
49          & COMPAS            & HFCN-C          & FCN             & 1,000             & original       & structured     & C             & N+W           \\
50          & Boston            & FCN-B           & FCN             & 102               & original       & structured     & R             & N+W \\
\hline \bottomrule
\multicolumn{9}{l}{\huge $\bullet$ where ``C'' and ``R'' represent classification/regression tasks. 
``N'', ``P'' and ``T'' represent normal training,}\\
\multicolumn{9}{l}{\huge poisoning and transfer learning scenarios.
``W'' and ``B'' represent white/black-box prioritization scenarios.}\\
\end{tabular} }
\end{table}

(3) \textbf{Various data types of test inputs}.
We mainly consider three data types, including
original test inputs, adversarial test inputs, and adaptive attacked test inputs.
For \textit{adversarial test inputs},
we perform three widely-used adversarial attacks 
(i.e.,
basic iterative method (BIM)~\cite{Kurakin2017AdversarialBIM},
Carlini \& Wagner (C\&W)~\cite{Carlini2017TowardsCW}
and FineFool~\cite{Chen2021FineFool})
to generate the same number of adversarial examples as the corresponding original test inputs for CIFAR10 and ImageNet,
respectively.
Then,
following previous works~\cite{Feng2020DeepGini,Wang2021PRIMA},
we construct an adversarial test input set for each of the two datasets under each adversarial attack through randomly selecting half of the original test inputs and half of the adversarial examples,
represented as ``+BIM'', ``+C\&W'', ``+FineFool''.
For \textit{adaptive attacked test inputs},
we follow the strategy introduced in Section~\ref{Intro} to produce them.
Regarding the surprise-based method~\cite{Kim2019Guiding},
we consider two objectives, 
flipping the labels of test inputs, 
and reducing their surprise
(i.e., the activation difference between test inputs and the training inputs),
which can be conducted by MAG-GAN~\cite{Chen2020MAGGAN}.
Regarding confidence-based methods~\cite{Feng2020DeepGini,Shen2020MCP},
we first flip the labels of test inputs based on BIM~\cite{Kurakin2017AdversarialBIM},
and then increase their highest probability value (up to 0.99 for CIFAR10 and 0.90 for ImageNet) by continuously adding perturbations.
Regarding the mutation-based method~\cite{Wang2021PRIMA}, 
we first break its ranking model by adding noise to the extracted features,
and then add perturbations to the test inputs based on back propagation~\cite{Amari1993Backpropagation} to generate noisy features.
Finally, 
we construct an adaptive attacked test input set for each of the two datasets under each adaptive attack through randomly selecting half of the original test inputs and half of the adaptive attacked examples,
represented as ``+AdapS'', ``+AdapC'', ``+AdapM''.
When crafting adversarial or adaptive attacked test inputs, 
we assume that the adversary knows the prioritization details.

Additionally,
regarding VGG19-AD model,
we implement three test input types respectively, i.e.,
the original one,
the patched test inputs (randomly blocking 10\% of pixels for each test input),
and the saturation-modified test inputs (modifying the intensities of saturation channel~\cite{Hossein2018SemanticHSV} for each test input).

(4) \textbf{Various structures of deep models}.
We employ CNN (ID: 1-41, 43),
LSTM (ID: 42, 44-46),
GCN (ID: 47) and 
FCN (ID: 48-50).
The number of layers ranges from 3 to 101 across all models.

(5) \textbf{Various training scenarios}.
We set up three training scenarios, including
normal training,
poisoning and 
transfer learning.
For the \textit{poisoning scenario},
we first craft a poisoned training set on FMNIST dataset,
denoted as ``FMNIST\_P'',
where the poisoning method is DeepPoison~\cite{Chen2021DeepPoison},
the source label and poisoned label are ``Pullover'' and ``Coat'', respectively.
For the \textit{transfer learning scenario},
we deploy the source domain as ImageNet with 1,000-class and the target domain as insect images or pet images with 2-class.

(6) \textbf{Various prioritization scenarios}.
We set up two prioritization scenarios, including 
white-box and black-box.
All details of the model and test inputs are available and used for prioritization in the \textit{white-box scenario} (ID: 1, 3-9, 11-17, 19-25, 27-33, 35-50),
while only model outputs and test inputs are available in the \textit{black-box scenario} (ID: 2, 10, 18, 26, 34).

{\renewcommand\arraystretch{0.82}
\begin{table}[t]
\caption{Application scope of prioritization methods.}
\label{tab:app_scope}
    \centering
    \resizebox{\linewidth}{!}{
    \begin{tabular}{l|cccc}
    \toprule \hline
        \textbf{Methods\quad} & \textbf{\quad Classification \quad} & \textbf{Regression} & \textbf{White-bok} & \textbf{\quad Black-box \quad} \\ \hline
        LSA & \checkmark & \checkmark & \checkmark & \tiny{\XSolid} \\
        DSA & \checkmark & \tiny{\XSolid} & \checkmark & \tiny{\XSolid} \\
        MCP & \checkmark & \tiny{\XSolid} & \checkmark & \checkmark \\
        DeepGini & \checkmark & \tiny{\XSolid} & \checkmark & \checkmark \\
        PRIMA & \checkmark & \checkmark & \checkmark & \checkmark\!\!\!$\backprime$ \\
        \textbf{CertPri} & \checkmark & \checkmark & \checkmark & \checkmark \\ \hline \bottomrule
    \end{tabular}
    } 
\end{table}
}

\subsection{Baselines}
We consider five baselines, i.e.,
likelihood-based surprise adequacy (LSA)~\cite{Kim2019Guiding},
distance-based surprise adequacy (DSA)~\cite{Kim2019Guiding},
multiple-boundary clustering and prioritization (MCP)~\cite{Shen2020MCP},
DeepGini~\cite{Feng2020DeepGini} and
PRIMA~\cite{Wang2021PRIMA}.
LSA and DSA are surprise-based methods.
MCP and DeepGini are efficient confidence-based methods for lightweight prioritization.
PRIMA is an effective mutation-based method with the SOTA performance.
The application scope of each method is shown in Table~\ref{tab:app_scope}, 
where ``\checkmark\!\!\!$\backprime$ '' means PRIMA can only perform input mutation in the black-box scenario.
Note that coverage-based methods have been shown to be significantly less effective~\cite{Feng2020DeepGini} and thus are omitted.
All baselines are configured according to the best performance setting reported in the respective papers.

\begin{table*}[t]
\centering
\caption{Overall comparison results across all subjects,
measured based on average RAUC and its improvement,
where ``C'' and ``R'' represent classification and regression,
``n/a'' means not applicable in theory.}
\label{tab:overallEffectiveness}
\resizebox{\linewidth}{!}{ \LARGE
\begin{tabular}{ll|lllll|rrrrr}
\toprule \hline
\multirow{2}{*}{\textbf{Tasks}} & \multirow{2}{*}{\textbf{Methods\quad}} & \multicolumn{5}{c|}{\textbf{Average $\pm$ {\large{Standard Deviation}} RAUC-}}                                                                                                                                              & \multicolumn{5}{c}{\textbf{Improvement of CertPri in RAUC-}}                             \\
                                &                                   & \multicolumn{1}{r}{\textbf{100}} & \multicolumn{1}{r}{\textbf{200}} & \multicolumn{1}{r}{\textbf{300}} & \multicolumn{1}{r}{\textbf{500}} & \multicolumn{1}{r|}{\textbf{All}} & \textbf{100} & \textbf{200} & \textbf{300} & \textbf{500} & \textbf{All} \\ \hline
\multirow{6}{*}{\textbf{C}}     & LSA                               & 0.4401$\pm$\large{0.2076}                & 0.3950$\pm$\large{0.2380}                & 0.3980$\pm$\large{0.2313}                & 0.3813$\pm$\large{0.2481}                & 0.5851$\pm$\large{0.1414}                 & 86.43\%      & 106.30\%     & 99.14\%      & 102.36\%     & 54.48\%      \\
                                & DSA                               & 0.4688$\pm$\large{0.2267}                & 0.4461$\pm$\large{0.2340}                & 0.4527$\pm$\large{0.2454}                & 0.4265$\pm$\large{0.2398}                & 0.5926$\pm$\large{0.1473}                 & 75.03\%      & 82.67\%      & 75.07\%      & 80.93\%      & 52.55\%      \\
                                & MCP                               & 0.4464$\pm$\large{0.2162}                & 0.4721$\pm$\large{0.2244}                & 0.4830$\pm$\large{0.2301}                & 0.4885$\pm$\large{0.2169}                & 0.6776$\pm$\large{0.1553}                 & 83.73\%      & 72.61\%      & 64.10\%      & 57.96\%      & 33.40\%      \\
                                & DeepGini                          & 0.6150$\pm$\large{0.2045}                & 0.5914$\pm$\large{0.2075}                & 0.5914$\pm$\large{0.2117}                & 0.5766$\pm$\large{0.2072}                & 0.7293$\pm$\large{0.1784}                 & 33.42\%      & 37.79\%      & 34.02\%      & 33.82\%      & 23.94\%      \\
                                & PRIMA                             & 0.6794$\pm$\large{0.2174}                & 0.6700$\pm$\large{0.2043}                & 0.6698$\pm$\large{0.2125}                & 0.6633$\pm$\large{0.2096}                & 0.7666$\pm$\large{0.1683}                 & 20.78\%      & 21.61\%      & 18.34\%      & 16.34\%      & 17.92\%      \\
                                & \textbf{CertPri}                  & \textbf{0.8205}$\pm$\large{0.1126}                & \textbf{0.8148}$\pm$\large{0.1011}                & \textbf{0.7926}$\pm$\large{0.1123}                & \textbf{0.7717}$\pm$\large{0.1205}                & \textbf{0.9040}$\pm$\large{0.0517}                 & n/a          & n/a          & n/a          & n/a           & n/a          \\ \cline{2-12}
\multirow{3}{*}{\textbf{R}}     & LSA                               & 0.4288$\pm$\large{0.1357}                & 0.3837$\pm$\large{0.0328}                & 0.4570$\pm$\large{0.0974}                & 0.4163$\pm$\large{0.0483}                & 0.6607$\pm$\large{0.0216}                 & 90.18\%      & 113.56\%     & 80.72\%      & 98.03\%      & 30.15\%      \\
                                & PRIMA                             & 0.7164$\pm$\large{0.1590}                & 0.6992$\pm$\large{0.1747}                & 0.6910$\pm$\large{0.1847}                & 0.6728$\pm$\large{0.2033}                & 0.7263$\pm$\large{0.1607}                 & 13.83\%      & 17.22\%      & 19.51\%      & 22.54\%      & 18.39\%      \\
                                & \textbf{CertPri}                  & \textbf{0.8154}$\pm$\large{0.0225}                & \textbf{0.8195}$\pm$\large{0.0161}                & \textbf{0.8258}$\pm$\large{0.0228}                & \textbf{0.8244}$\pm$\large{0.0188}                & \textbf{0.8599}$\pm$\large{0.0150}                 & n/a          & n/a          & n/a          & n/a           & n/a   \\ \hline \bottomrule      
\end{tabular}
}
\end{table*}

\subsection{Measurements}
We investigate CertPri's prioritization performance from five aspects, 
including 
prioritization \textit{effectiveness}, \textit{efficiency}, \textit{robustness}, \textit{generalizability} and \textit{guidance}.

(1)~We evaluate the effectiveness of CertPri through the ratio of the area under curve (RAUC), 
which transforms the prioritization result to a curve~\cite{Wang2021PRIMA},
defined as follows for classification tasks:
\begin{equation} \footnotesize
    \!\!\!\! {\rm RAUC} \! = \! \frac{ \sum_{i=1}^{N} n_{i} }{ N\!\!\times\!\! N'\! + \frac{ N'\! - N'^{2} }{2} }, 
    {\rm ~where}~n_{i}\! = \!\left\{
    \begin{array}{l}
    \!\!\!n_{i\textrm{-}1} \!+\! 1,~c(\bm{x}_{i}) {\rm ~is~correct}  \\
    \!\!\!n_{i\textrm{-}1},~{\rm otherwise}
    \end{array}
    \right.\!\!\!\!,
\end{equation}
where $N$ and $N'$ are the number of prioritized test inputs and bug-revealing inputs,
$i\ge1$ and $n_{0}\!=\!0$.
In other words, 
the numerator and denominator represent
the area under curve of the prioritization method and the ideal prioritization, 
respectively.

For regression tasks, 
RAUC is calculated based on the mean-square error (MSE) of prediction,
as follows:
\begin{equation} \footnotesize
    {\rm RAUC} = \frac{ \sum_{i=1}^{N} m_{i} }{ \sum_{j=1}^{N} M_{j} },~
    {\rm where} \left\{
    \begin{array}{l}
         m_{i} = m_{i\textrm{-}1} + {\rm MSE} \left(f^{R}(\bm{x}_{i}) \right)  \\
         M_{j} \!=\! M_{j\textrm{-}1} + {\rm MSE} \left(f^{R}(\bm{x}_{j}) \right) 
    \end{array}
    \right.\!\!\!\!,
\end{equation}
where $m_{i}$ and $M_{j}$ represent the accumulated MSE between the predicted results and the ground-truth of the prioritization method and the ideal prioritization, 
respectively.
$i,j\ge1$, $m_{0}\!=\!M_{0}\!=\!0$.
Moreover,
we follow the setup of Wang \textit{et al.}~\cite{Wang2021PRIMA}, 
using RAUC-100, RAUC-200, RAUC-300, RAUC-500 and RAUC-all as fine-grained metrics, 
i.e. $N$=100, 200, 300, 500 and all.
Larger RAUC is better.

(2)~We evaluate the prioritization efficiency of CertPri through prioritization speed,
i.e., the time cost of prioritize 1,000 test inputs (\#seconds/1,000 Inputs). Smaller is better.

(3)~We evaluate the prioritization robustness of CertPri through RAUC stability,
denoted as RobR.
Larger RobR is better,
as follows:
\begin{equation} \footnotesize
    \rm RobR = \frac{~~RAUC\textrm{-}all~ of~ adaptive~ attacked~ test~ inputs~~}{RAUC\textrm{-}all~ of~ original~ test~ inputs} \times 100\%.
    \label{equ:RobR}
\end{equation}

(4)~We evaluate the prioritization generalizability of CertPri based on reward value,
denoted as GenRew,
as follows:
\begin{equation} \small
    {\rm GenRew} = \frac{1}{N_{rep} \times N_{sub}}\sum_{i=1}^{N_{rep}}\sum_{j=1}^{N_{sub}} \frac{n_{pm} - k_{i,j} + 1}{n_{pm}},
    \label{equ:GenRew}
\end{equation}
where $N_{sub}$ represents the number of selected subjects,
$N_{rep}\!\!=\!\!5$ and $n_{pm} \!\!=\!\! 6$ are the number of repetitions and prioritization methods in our experiments,
respectively.
$k_{i,j}\in \{ 1,2,...,n_{pm} \}$ represents that 
the method ranks in the $k$-th position in descending order of RAUC-all for the $j$-th subject at the $i$-th experimental repetition.
GenRew $\in [\frac{1}{n_{pm}} , 1]$.
Lager GenRew is better.

(5) We evaluate the prioritization guidance from two aspects:
performance and robustness improvements for DNNs.
The former subtracts the accuracy of original model from that of retrained model,
while the latter subtracts the attack success rate~\cite{Chen2020MAGGAN} of retrained model from that of original model.

\subsection{Implementation Details}
To fairly study the performance of the baselines and CertPri, 
our experiments have the following settings.
\textbf{(1)~Hyperparameter settings} based on the double-minimum strategy: 
we conduct a preliminary study based on a small dataset, 
and find that $N_{b}$>3, $N_{rsb}$>5 and $0.02\max(\bm{x})$<$\mathcal{R}$<$0.05\max(\bm{x})$ for Algorithm 1 are effective in general.
To guarantee CertPri's effectiveness, 
we follow the double-minimum strategy, 
i.e., $N_{b}$=6, $N_{rsb}$=10, $\mathcal{R}$=0.04$\max(\bm{x})$, 
$p$=2 for Algorithm 1.
\textbf{(2)~Model training}: 
we download and use pretrained models on ImageNet. 
For other datasets, 
we train appropriate models as follows. 
The learning rate ranges from 1E-04 to 1E-02, 
the optimizer is Adam, 
training:validation:test = 7:1:2, 
and an early stop strategy is used to avoid overfitting.
\textbf{(3)~Preprocessing}:
we fill in missing data points based on the mean value.
There is no specific mutation strategy provided in PRIMA~\cite{Wang2021PRIMA} for speech, signal, graph data forms and GCN model, 
thus we derive it from the existing mutation operations.
\textbf{(4)~Data recording}:
we repeat the experiment 5 times and record about 9,000 raw data. 

We conduct all the experiments on a server with one Intel i7-7700K CPU running at 4.20 GHz, 
64 GB DDR4 memory, 
4 TB HDD and one TITAN Xp 12 GB GPU card.

\section{Experimental Results and Analysis\label{Experiment}}

We evaluate CertPri through answering the following research questions (RQ).
\\
\textbf{RQ1.~Effectiveness}: How \textit{effective} is CertPri?
\\
\textbf{RQ2.~Efficiency}: How \textit{efficient} is CertPri.?
\\
\textbf{RQ3.~Robustness}: How \textit{robust} is CertPri?
\\
\textbf{RQ4.~Generalizability}: How \textit{generic} is CertPri?
\\
\textbf{RQ5.~Guidance}: Can CertPri \textit{guide} the retraining of DNNs?

\subsection{Effectiveness (RQ1)}
\begin{center}
\fcolorbox{black}{gray!20}{\parbox{0.97\linewidth}
    {
        How \textit{effective} is CertPri in prioritizing test inputs?
    }
}
\end{center}

When reporting the results,
we focus on the effectiveness of the following aspects: 
overall, 
data forms, 
data types, 
training scenarios, and 
prioritization scenarios.
The evaluation results are shown in Tables~\ref{tab:overallEffectiveness}, \ref{tab:T-test}, \ref{tab:DataFormsEffectiveness}, \ref{tab:DataTypesEffectiveness}, \ref{tab:TraSceEffectiveness} and \ref{tab:PriSceEffectiveness}.

\begin{table}[ht]
\centering
\caption{The p-value of T-test for average RAUC between CertPri and each baseline.}
\label{tab:T-test}
\resizebox{\linewidth}{!}{ \large
\begin{tabular}{l|rrrrr}
\toprule \hline
         & \multicolumn{5}{c}{\textbf{CertPri}}                             \\ \cline{2-6}
         & RAUC-100 & RAUC-200 & RAUC-300 & RAUC-500 & RAUC-all \\ \hline
LSA      & 1.79E-19 & 2.61E-18 & 6.40E-19 & 1.21E-17 & 4.99E-21 \\
DSA      & 8.52E-14 & 1.70E-14 & 5.79E-14 & 2.02E-15 & 4.39E-18 \\
MCP      & 7.43E-15 & 4.58E-14 & 8.68E-14 & 2.52E-14 & 3.39E-12 \\
DeepGini & 4.01E-09 & 5.02E-10 & 2.56E-09 & 2.19E-10 & 2.48E-08 \\
PRIMA    & 4.24E-06 & 3.21E-07 & 1.39E-06 & 2.22E-06 & 7.53E-08 \\
\hline \bottomrule
\end{tabular}
}
\end{table}

\begin{table}[ht]
\centering
\caption{Comparison on various data forms of test inputs across all subjects.}
\label{tab:DataFormsEffectiveness}
\resizebox{\linewidth}{!}{\footnotesize
\begin{tabular}{ll|rrrrr}
\toprule \hline
\multirow{2}{*}{\textbf{Data Forms}} & \multirow{2}{*}{\textbf{Methods\quad}} & \multicolumn{5}{c}{\textbf{Average RAUC-}}                                              \\
                                       &                                   & \textbf{100}    & \textbf{200}    & \textbf{300}    & \textbf{500}    & \textbf{All}    \\ \hline
\multirow{3}{*}{Image}                 & DeepGini                          & 0.6286          & 0.5919          & 0.5842          & 0.5709          & 0.7047          \\
                                       & PRIMA                             & 0.7076          & 0.6952          & 0.6785          & 0.6713          & 0.7477          \\
                                       & \textbf{CertPri}                  & \textbf{0.8449} & \textbf{0.8398} & \textbf{0.8128} & \textbf{0.7890} & \textbf{0.9060} \\ \cline{2-7}
\multirow{3}{*}{Text}                  & DeepGini                          & 0.5312          & 0.5766          & 0.6213          & 0.6481          & 0.8516          \\
                                       & PRIMA                             & 0.6438          & 0.6391          & 0.6533          & 0.7087          & 0.8474          \\
                                       & \textbf{CertPri}                  & \textbf{0.7182} & \textbf{0.7099} & \textbf{0.7315} & \textbf{0.7607} & \textbf{0.9145} \\ \cline{2-7}
\multirow{3}{*}{Speech}                & DeepGini                          & 0.6251          & 0.6307          & 0.7924          & n/a             & 0.8584          \\
                                       & PRIMA                             & 0.4202          & 0.4966          & 0.8436          & n/a             & \textbf{0.8806} \\
                                       & \textbf{CertPri}                  & \textbf{0.7437} & \textbf{0.7676} & \textbf{0.8596} & n/a             & 0.8535          \\ \cline{2-7}
\multirow{3}{*}{Signal}                & DeepGini                          & n/a             & n/a             & n/a             & n/a             & n/a             \\
                                       & PRIMA                             & 0.7233          & 0.7068          & 0.7296          & n/a             & 0.7521          \\
                                       & \textbf{CertPri}                  & \textbf{0.8182} & \textbf{0.7980} & \textbf{0.8192} & n/a             & \textbf{0.8444} \\ \cline{2-7}
\multirow{3}{*}{Graph}                 & DeepGini                          & 0.6262          & 0.6059          & 0.5825          & 0.5677          & \textbf{0.8482} \\
                                       & PRIMA                             & 0.4193          & 0.4516          & 0.5072          & 0.4944          & 0.6541          \\
                                       & \textbf{CertPri}                  & \textbf{0.7503} & \textbf{0.7367} & \textbf{0.7059} & \textbf{0.6941} & 0.8417          \\ \cline{2-7}
\multirow{3}{*}{Structured}            & DeepGini                          & 0.5281          & 0.5839          & 0.5617          & 0.5390          & 0.8044          \\
                                       & PRIMA                             & 0.6582          & 0.5922          & 0.5850          & 0.5513          & 0.7846          \\
                                       & \textbf{CertPri}                  & \textbf{0.6719} & \textbf{0.6199} & \textbf{0.6010} & \textbf{0.5986} & \textbf{0.8317} \\
                                       \hline \bottomrule
\end{tabular}
}
\end{table}

Implementation details for effectiveness evaluation.
(1) We present the overall comparison results in Table~\ref{tab:overallEffectiveness}.
Since DSA, MCP and DeepGini cannot directly apply to regression tasks, 
we separately present results on both tasks.
(2) We conduct a preliminary T-test about RAUC across all subjects,
as shown in Table~\ref{tab:T-test}. 
(3) Since DeepGini and PRIMA show significantly better performance than other baselines, 
we mainly compare the results between CertPri and these two in Tables~\ref{tab:DataFormsEffectiveness}, \ref{tab:DataTypesEffectiveness}, \ref{tab:TraSceEffectiveness} and \ref{tab:PriSceEffectiveness}.

\textbf{Overall effectiveness}.
CertPri finds a better permutation of test inputs than baselines,
i.e.,
identifying bug-revealing inputs earlier,
which significantly improves prioritization.
For instance,
in Table~\ref{tab:overallEffectiveness},
all average RAUC values of CertPri are the highest compared with baselines for various tasks.
More specifically, 
the average RAUC value of CertPri is 1.50 times and 1.42 times that of baselines for classification and regression tasks, respectively.
Additionally, 
CertPri improves the prioritization effect by 55.39\% and 50.41\% for classification and regression tasks, respectively.
From Table~\ref{tab:T-test}, 
we can see that the p-values of all RAUC metrics are small enough, 
which demonstrates that CertPri significantly outperforms all baselines.
The outstanding performance of CertPri is mainly because 
it takes formal guarantees into account when identifying bug-revealing inputs while baselines only prioritize empirically.
Therefore, 
CertPri shows a stable overall prioritization effect without being affected by interfering factors
(e.g., various tasks, various data types, etc).

\begin{table}[ht]
\centering
\caption{Comparison on various data types of test inputs across the image subjects (ID: 3-8, 11-16, 19-24, 27-32).}
\label{tab:DataTypesEffectiveness}
\resizebox{\linewidth}{!}{
\begin{tabular}{ll|rrrrr}
\toprule \hline
\multirow{2}{*}{\textbf{Data Types}}                                         & \multirow{2}{*}{\textbf{Methods\quad}} & \multicolumn{5}{c}{\textbf{Average RAUC-}}                                              \\
                                                                             &                                   & \textbf{100}    & \textbf{200}    & \textbf{300}    & \textbf{500}    & \textbf{All}    \\ \hline
\multirow{3}{*}{Adversarial}                                                 & DeepGini                          & 0.8187          & 0.7960          & 0.7930          & 0.7592          & 0.7790          \\
                                                                             & PRIMA                             & \textbf{0.9379} & 0.9212          & \textbf{0.9224} & 0.8963          & 0.8510          \\
                                                                             & \textbf{CertPri}                  & 0.9219          & \textbf{0.9306} & 0.9184          & \textbf{0.9015} & \textbf{0.9441} \\ \cline{2-7}
\multirow{3}{*}{\begin{tabular}[c]{@{}l@{}}Adaptive\\ Attacked\end{tabular}} & DeepGini                          & 0.4220          & 0.3719          & 0.3609          & 0.3575          & 0.5359          \\
                                                                             & PRIMA                             & 0.5000          & 0.4951          & 0.4581          & 0.4685          & 0.5667          \\
                                                                             & \textbf{CertPri}                  & \textbf{0.8183} & \textbf{0.7989} & \textbf{0.7517} & \textbf{0.7006} & \textbf{0.8827} \\ \hline \bottomrule
\end{tabular}
}
\end{table}

\begin{table}[ht]
\centering
\caption{Comparison on various training scenarios across the image subjects (ID: 38-40).}
\label{tab:TraSceEffectiveness}
\resizebox{\linewidth}{!}{ \small
\begin{tabular}{ll|rrrrr}
\toprule \hline
\multirow{2}{*}{\textbf{\begin{tabular}[c]{@{}l@{}}Training\\ Scenarios\end{tabular}}} & \multirow{2}{*}{\textbf{Methods\quad}} & \multicolumn{5}{c}{\textbf{Average RAUC-}}                                              \\
                                                                                       &                                   & \textbf{100}    & \textbf{200}    & \textbf{300}    & \textbf{500}    & \textbf{All}    \\ \hline
\multirow{3}{*}{Poisoning}                                                             & DeepGini                          & 0.5021          & 0.4849          & 0.4571          & 0.4921          & 0.5457          \\
                                                                                       & PRIMA                             & \textbf{0.9949} & 0.9855          & 0.9834          & \textbf{0.9815} & 0.9785          \\
                                                                                       & \textbf{CertPri}                  & 0.9769          & \textbf{0.9908} & \textbf{0.9899} & 0.9790          & \textbf{0.9889} \\ \cline{2-7}
\multirow{3}{*}{\begin{tabular}[c]{@{}l@{}}Transfer\\ Learning\end{tabular}}           & DeepGini                          & 0.7787          & 0.7017          & 0.7198          & 0.7320          & 0.8580          \\
                                                                                       & PRIMA                             & \textbf{0.8790} & 0.8221          & 0.8075          & 0.8073          & 0.9073          \\
                                                                                       & \textbf{CertPri}                  & 0.8694          & \textbf{0.8274} & \textbf{0.8280} & \textbf{0.8470} & \textbf{0.9532} \\ \hline \bottomrule
\end{tabular}
}
\end{table}

\begin{table}[ht]
\centering
\caption{Comparison on various prioritization scenarios across the image subjects (ID: 1-36).}
\label{tab:PriSceEffectiveness}
\resizebox{\linewidth}{!}{ \small
\begin{tabular}{ll|rrrrr}
\toprule \hline
\multirow{2}{*}{\textbf{\begin{tabular}[c]{@{}l@{}}Prioritization\\ Scenarios\end{tabular}}} & \multirow{2}{*}{\textbf{Methods}} & \multicolumn{5}{c}{\textbf{Average RAUC-}}                                              \\
                                                                                             &                                   & \textbf{100}    & \textbf{200}    & \textbf{300}    & \textbf{500}    & \textbf{All}    \\ \hline
\multirow{3}{*}{White-box}                                                                   & DeepGini                          & 0.6246          & 0.5886          & 0.5794          & 0.5603          & 0.6774          \\
                                                                                             & PRIMA                             & 0.7380          & 0.7260          & 0.7091          & 0.6959          & 0.7422          \\
                                                                                             & \textbf{CertPri}                  & \textbf{0.8563} & \textbf{0.8514} & \textbf{0.8214} & \textbf{0.7909} & \textbf{0.9039} \\ \cline{2-7}
\multirow{3}{*}{Black-box}                                                                   & DeepGini                          & 0.6499          & 0.6164          & 0.5939          & 0.5724          & 0.7975          \\
                                                                                             & PRIMA                             & 0.4220          & 0.4471          & 0.4063          & 0.3964          & 0.5932          \\
                                                                                             & \textbf{CertPri}                  & \textbf{0.8108} & \textbf{0.7896} & \textbf{0.7463} & \textbf{0.7227} & \textbf{0.8688} \\ \hline \bottomrule
\end{tabular}
}
\end{table}

\textbf{Effectiveness on various data forms of test inputs}.
CertPri outperforms all baselines on all six data forms in terms of almost all RAUC values,
especially for unstructured data forms (i.e., image, text, etc).
For instance,
in Table~\ref{tab:DataFormsEffectiveness}, 
almost all RAUC values of CertPri are the highest, 
except RAUC-all on speech and graph data forms.
We investigate their model feature space and 
find that their decision boundaries are smoother than others, 
which causes gradient vanishing.
Therefore,
we extend the sampling radius $\mathcal{R} = 0.05\max(\bm{x})$, 
which facilitates maximum gradient norm estimation based on GEVT. 
After radius extension, 
CertPri realizes the highest average RAUC-all, 
improving to 0.9127 and 0.8817 for speech and graph data forms, respectively.
Additionally, 
the average RAUC of CertPri is 1.13$\sim$1.30 times that of baselines for unstructured data forms,
but only 1.07 times for structured data.
We speculate that the gradient vanishes during back propagation due to the sparse coding of structured data.
Gradient vanishing is beyond the scope of this paper, 
but it can be improved by 
batch normalization~\cite{Ioffe2015BatchBatchNormalization} and
non-saturating activation function~\cite{Konstantin2019ReLU}.

\textbf{Effectiveness on various data types of test inputs}.
CertPri largely outperforms all baselines against adaptive attacks in terms of all average RAUC values,
while approaching the SOTA baseline (i.e., PRIMA) against adversarial attacks.
For instance,
in Table~\ref{tab:DataTypesEffectiveness}, 
the average RAUC values of CertPri against adaptive attacks range from 0.7006 to 0.8827 with average improvements of 64.73\%$\sim$114.85\% compared with DeepGini and
49.55\%$\sim$64.11\% compared with PRIMA,
respectively.
This is mainly because we provide a formal guarantee of movement costs in feature space,
which cannot be an objective of adaptive attacks.
Besides,
the average RAUC-100 and RAUC-300 gaps between PRIMA and CertPri are both less than 0.02,
which is acceptable and does not hinder its practical application in various data types.

\textbf{Effectiveness on various training scenarios}.
CertPri outperforms DeepGini and shows competitive performance with PRIMA in both training scenarios.
For instance,
in Table~\ref{tab:TraSceEffectiveness},
the average RAUC values of CertPri range from 0.9769 to 0.9908 with average improvements of 81.20\%$\sim$116.56\% compared with DeepGini in the poisoning scenario,
and range from 0.8274 to 0.9532 with average improvements of 11.10\%$\sim$17.92\% compared with DeepGini in transfer learning.
Besides,
the average RAUC-100 and RAUC-500 gaps between PRIMA and CertPri are both less than 0.02.
We speculate that the purity assumption of DeepGini is not tenable in the two training scenarios~\cite{Wang2021PRIMA}, 
whereas CertPri and PRIMA facilitate their prioritization based on the movement cost and mutation perspectives, 
respectively.

\textbf{Effectiveness on various prioritization scenarios}.
CertPri largely outperforms all baselines in terms of all RAUC values in both prioritization scenarios,
which is beneficial to identify bug-revealing inputs in software engineering testing with privacy requirements.
For instance,
in Table~\ref{tab:PriSceEffectiveness},
the average RAUC values of CertPri range from 0.7909 to 0.9039 with average improvements of 13.66\%$\sim$44.65\% compared with baselines in the white-box scenario,
and range from 0.7227 to 0.8688 with average improvements of 8.94\%$\sim$92.12\% compared with baselines in the black-box scenario.
The outstanding performance of CertPri is mainly because it 
leverages the Weibull distribution to determine the exact maximum gradient norm in the white-box scenario,
and adopts gradient estimation to satisfy the black-box scenario.

\begin{center}
\fcolorbox{black}{white!20}{\parbox{0.97\linewidth}
    {
        \emph{\textbf{Answer to RQ1}}:
        CertPri outperforms baselines in two aspects in terms of effectiveness:
        (1)~\textit{overall}-it significantly improves 81.84\%, 47.48\% and 18.65\% RAUC values on average compared with surprise-based, confidence-based and mutation-based baselines, respectively;
        (2)~\textit{specific}-it improves 20.66\%, 43.39\%, 28.96\% and 38.88\% RAUC values on average compared with baselines (i.e., DeepGini and PRIMA) for various data forms, data types, training scenarios and prioritization scenarios, respectively.
    }
}
\end{center}

\subsection{Efficiency (RQ2)}
\begin{center}
\fcolorbox{black}{gray!20}{\parbox{0.97\linewidth}
    {
        How \textit{efficient} is CertPri in prioritizing test inputs?
    }
}
\end{center}

When answering this question, 
we refer to the prioritization time costs in the white-box scenarios with image data form (i.e., ID: 1-36).
The evaluation results are shown in Table~\ref{tab:Efficiency},
where the time cost of PRIMA includes input mutation, model mutation and ranking model training.
Here we have the following observation.

\textbf{Prioritization efficiency}.
CertPri prioritizes test inputs more efficiently than mutation-based methods and is competitive with confidence-based methods,
which meets the rapidity requirements of software engineering testing. 
For instance,
in Table~\ref{tab:Efficiency},
the efficiency of CertPri is 41.17 times and 52.86 times that of DSA and PRIMA, 
respectively.
This is mainly because CertPri only needs to perform gradient computation based on back propagation and extreme value estimation, 
both of which are lightweight operations.
Besides,
in Table~\ref{tab:Efficiency},
the time cost of CertPri is 1.25 times and 2.94 times more than that of MCP and DeepGini on average, 
respectively.
The reason is that CertPri involves iterative operations in the extreme value estimation, 
which increases time costs.
Note that CertPri adopts a double-minimum strategy to ensure its effectiveness, 
which leaves room for efficiency improvements. 
We can slightly reduce the $N_{b}$, $N_{rsb}$, $\mathcal{R}$ values in \textbf{Algorithm 1} to further improve its efficiency without loss of effectiveness.

\begin{center}
\fcolorbox{black}{white!20}{\parbox{0.97\linewidth}
    {
        \emph{\textbf{Answer to RQ2}}:
        CertPri is more efficient in prioritization speed - it prioritizes test inputs with an average speedup of 51.86 times compared with SOTA method (i.e., PRIMA).
    }
}
\end{center}

\begin{table}[ht]
\centering
\caption{Efficiency comparison across the image subjects (ID: 1-36), measured by ``\#seconds/1,000 Inputs''.}
\label{tab:Efficiency}
\resizebox{\linewidth}{!}{\Huge
\begin{tabular}{ll|r|rrrrrr}
\toprule \hline
\multirow{2}{*}{\textbf{Datasets}} & \multirow{2}{*}{\textbf{Models}} & \multirow{2}{*}{\textbf{\begin{tabular}[c]{@{}r@{}}Model\\Weights \end{tabular}}} & \multicolumn{6}{c}{\textbf{Time Costs}}                                    \\ \cline{4-9}
                                   &                                  &                                                                                      & LSA      & DSA      & MCP      & DeepGini & PRIMA    & \textbf{CertPri} \\ \hline
\multirow{2}{*}{CIFAR10}           & ResNet50                         & 2.58E+07                                                                             & 1.78E+01 & 1.23E+03 & 1.24E+01 & 6.88E+00 & 1.65E+03 & 2.93E+01         \\
                                   & VGG16                            & 1.54E+07                                                                             & 1.34E+01 & 1.20E+03 & 7.55E+00 & 4.34E+00 & 1.29E+03 & 1.68E+01         \\ \cline{2-9}
\multirow{2}{*}{ImageNet}          & ResNet101                        & 4.47E+07                                                                             & 2.15E+02 & 1.49E+04 & 1.51E+02 & 8.34E+01 & 1.60E+04 & 3.55E+02         \\
                                   & VGG19                            & 1.44E+08                                                                             & 8.73E+02 & 1.72E+04 & 4.94E+02 & 2.84E+02 & n/a      & 1.10E+03         \\ \cline{2-9}
DrivingSA                          & VGG19-AD                         & 7.04E+07                                                                             & 3.67E+02 & 1.60E+04 & 2.07E+02 & 1.19E+02 & 1.53E+04 & 4.61E+02    \\ \hline \bottomrule      
\end{tabular}
}
\end{table}

\subsection{Robustness (RQ3)}
\begin{center}
\fcolorbox{black}{gray!20}{\parbox{0.97\linewidth}
    {
        How \textit{robust} is CertPri against adaptive attacks based on its certifiability?
    }
}
\end{center}

\begin{figure}[t]
    \centering
    \includegraphics[width=0.6\linewidth]{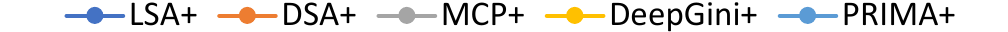} \\
    \subfigure[+AdapS (ID: 22)]{
        \includegraphics[width=0.29\linewidth]{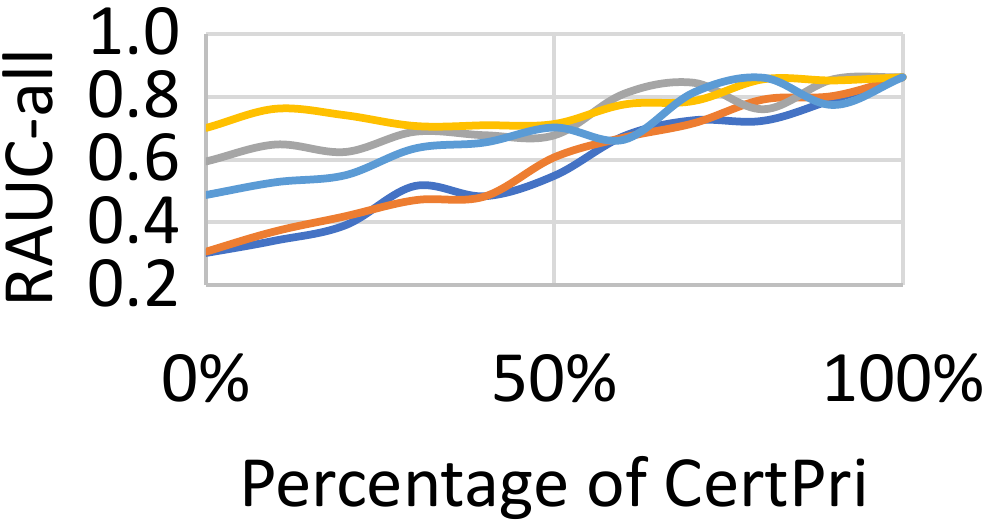}
        } 
    \subfigure[+AdapC (ID: 23)]{
        \includegraphics[width=0.29\linewidth]{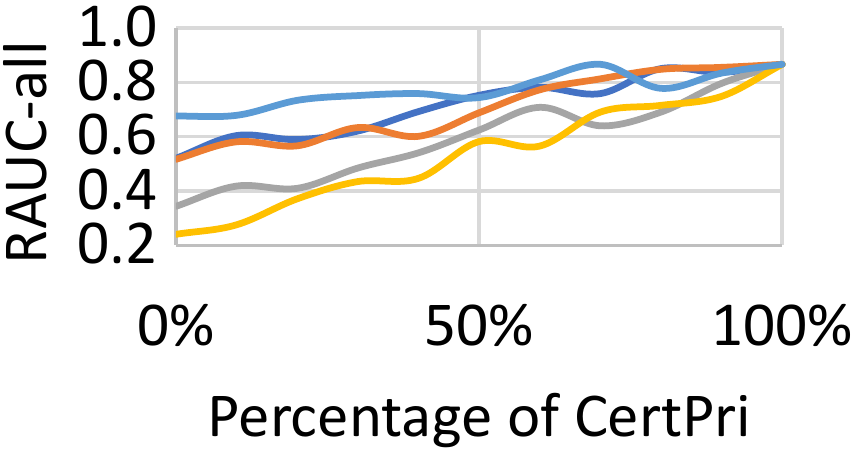}
        }
    \subfigure[+AdapM (ID: 24)]{
        \includegraphics[width=0.29\linewidth]{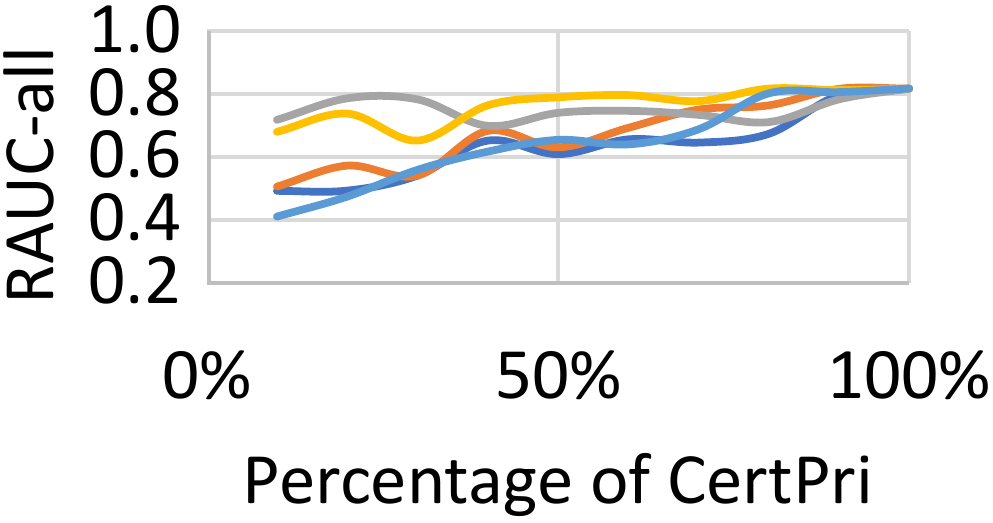}
        }
    \caption{The promotion results of CertPri for baselines through combining it with each baseline.}
    \label{fig:RobEnsemble}
\end{figure}

When reporting the results, 
we focus on the following aspects:
the robustness against adaptive attacks and the utility of robustness.
The evaluation results are shown in Table~\ref{tab:robustness} and Figure~\ref{fig:RobEnsemble}.

Implementation details for robustness evaluation.
(1)~To measure CertPri's robustness, 
we refer to the variation of RAUC values between original test inputs (ID: 1, 9, 17, 25) and adaptive attacked test inputs (ID: 6-8, 14-16, 22-24, 30-32),
i.e., RobR,
as shown in Table~\ref{tab:robustness}.
(2)~To demonstrate the robustness utility,
taking adaptive attacks on ImageNet dataset as an example (ID: 22-24),
we combine CertPri with each baseline to show the promotion effect of CertPri on baselines, 
as shown in Figure~\ref{fig:RobEnsemble},
where \textit{x}-axis represents the percentage of CertPri components added to baselines.

\begin{table}[t]
\centering
\caption{Robustness comparison across the image subjects (ID: 1, 6-9, 14-17, 22-25, 30-32), measured by RobR.}
\label{tab:robustness}
\resizebox{\linewidth}{!}{
\begin{tabular}{l|rrrrr}
\toprule \hline
\multicolumn{1}{c|}{\multirow{2}{*}{\textbf{Methods}}} & \multicolumn{5}{c}{\textbf{Prioritization Robustness in RAUC-}}                                                       \\
\multicolumn{1}{c|}{}                                  & \textbf{100}      & \textbf{200}      & \textbf{300}      & \textbf{500}     & \textbf{All}      \\
\hline
LSA                                                   & 66.58\%           & 52.20\%           & 65.67\%           & 50.24\%          & 68.77\%           \\
DSA                                                   & 66.17\%           & 65.16\%           & 68.18\%           & 66.46\%          & 71.28\%           \\
MCP                                                   & 53.92\%           & 55.79\%           & 63.61\%           & 54.73\%          & 68.92\%           \\
DeepGini\qquad                                              & 64.94\%           & 60.36\%           & 60.88\%           & 62.38\%          & 67.25\%           \\
PRIMA                                                 & 63.18\%           & 65.31\%           & 61.58\%           & 67.75\%          & 64.63\%           \\
\textbf{CertPri}                                      & \quad \textbf{101.31\%} & \quad \textbf{100.90\%} & \quad \textbf{103.60\%} & \quad \textbf{99.93\%} & \quad \textbf{100.77\%} \\
\hline \bottomrule 
\end{tabular}
}
\end{table}

\textbf{Prioritization robustness}.
In all cases,
CertPri always performs the most robust prioritization results against various adaptive attacks,
which facilitates stable identification of bug-revealing inputs.
For instance,
in Table~\ref{tab:robustness},
the average RobR values of CertPri against adaptive attacks range from 99.93\% to 103.60\%,
which is 1.50$\sim$1.71 times that of baselines.
The outstanding performance of CertPri is mainly because it
simplifies the prioritization task into a lower bound measure of the movement cost,
which has been formally guaranteed in Section~\ref{FormalGuarantees}.

\textbf{Utility of robustness}.
CertPri is not only immune to various adaptive attacks, 
but also facilitates the robustness of baselines through weighted combinations.
For instance,
in Figure~\ref{fig:RobEnsemble},
all curves show an upward trend after combining with CertPri and are always higher than their initial value without CertPri.
Furthermore, 
we speculate that for an empirical prioritization method that outperforms CertPri in general, 
it can be combined with CertPri to improve robustness against adaptive attacks.

\begin{center}
\fcolorbox{black}{white!20}{\parbox{0.97\linewidth}
    {
        \emph{\textbf{Answer to RQ3}}:
        CertPri largely outperforms baselines in terms of robust prioritization with average robustness improvements of 41.37\%$\sim$98.91\%.
        Besides, 
        its robustness can be leveraged to facilitate other methods.
    }
}
\end{center}

\subsection{Generalizability (RQ4)}
\begin{center}
\fcolorbox{black}{gray!20}{\parbox{0.97\linewidth}
    {
        How \textit{generic} is CertPri in prioritizing test inputs?
    }
}
\end{center}

\begin{figure}
    \centering
    \subfigure[GenRew of RAUC-300]{
        \includegraphics[width=0.38\linewidth]{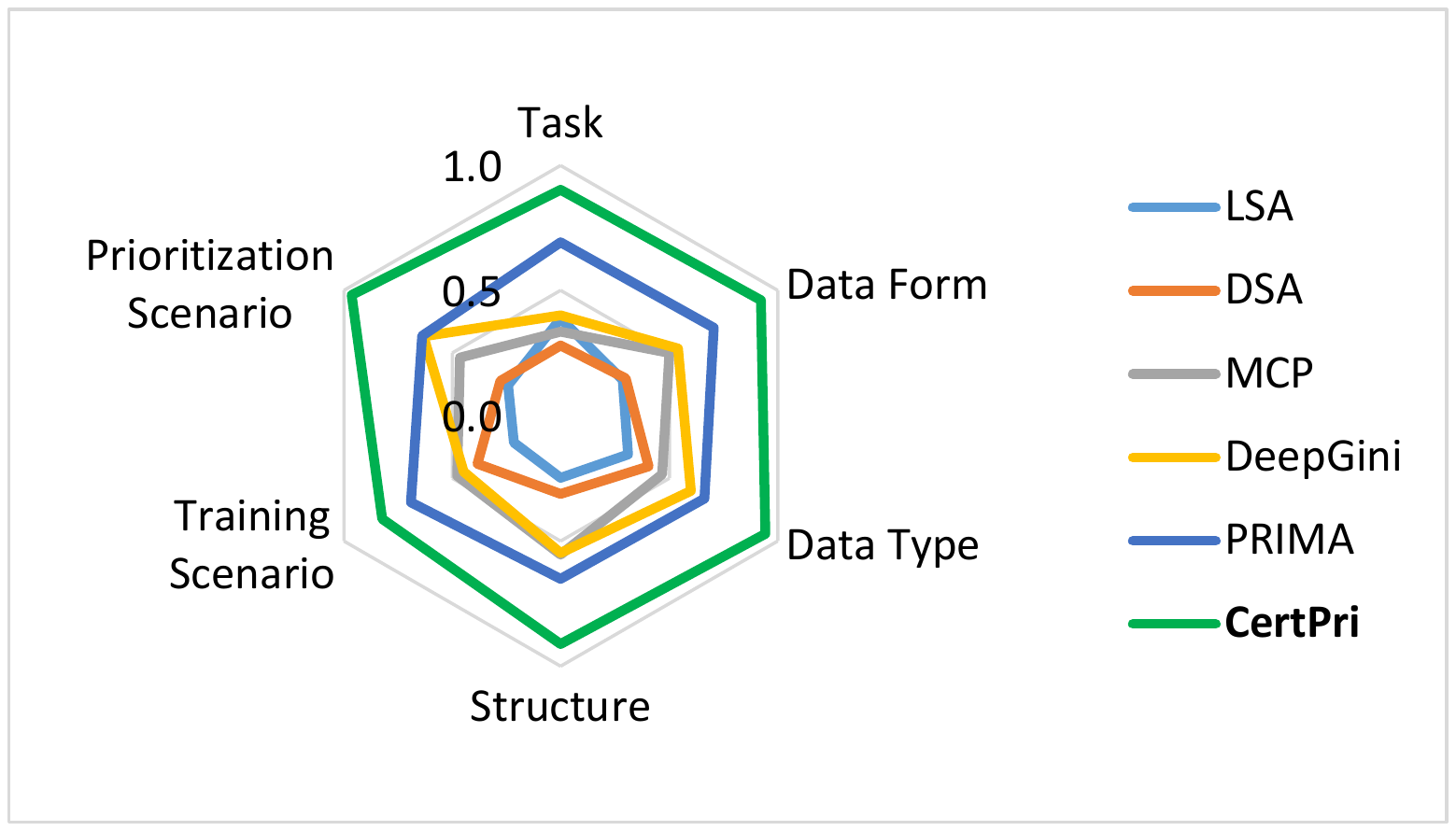}
    } 
    \subfigure[GenRew of RAUC-all]{
        \includegraphics[width=0.38\linewidth]{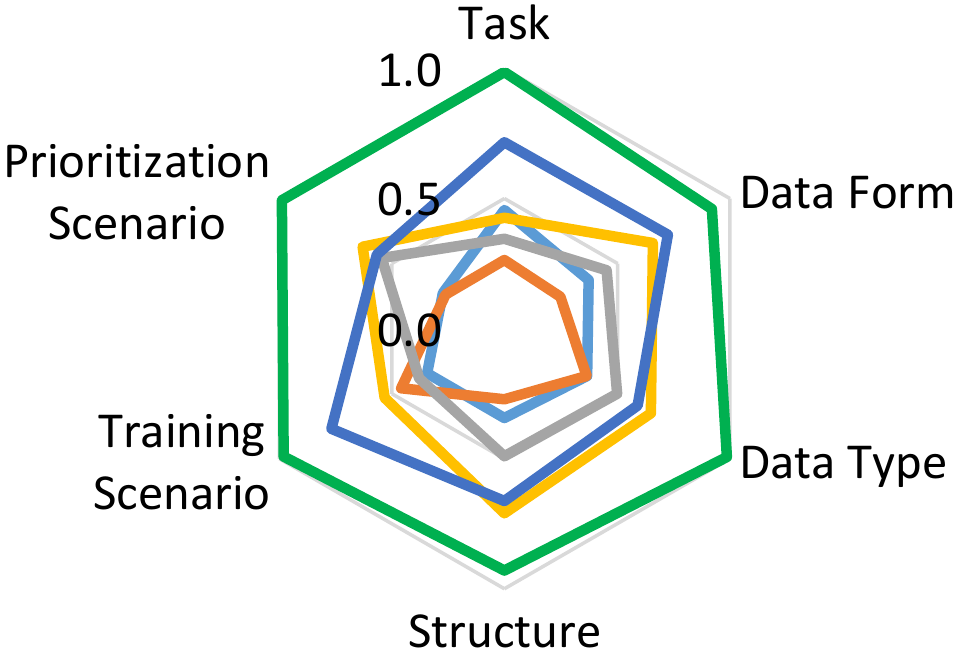}
    } 
    \includegraphics[width=0.11\linewidth]{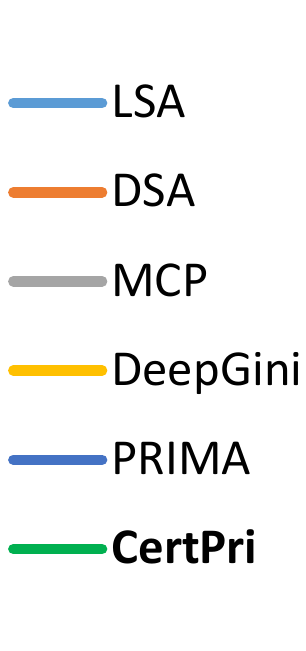}
    \caption{Generalizability comparison for six dimensions across all subjects, measured by GenRew.}
    \label{fig:generalizabilityRadar}
\end{figure}

When reporting the generalizability, 
we focus on the ranking of RAUC values for different methods in each subject.
The evaluation results are illustrated as radar charts in Figure~\ref{fig:generalizabilityRadar}.

Implementation details for generalizability evaluation.
(1)~Calculate the GenRew value of each dimension separately.
Take the task dimension as an example, 
which includes classification and regression.
First, 
we select all subjects belonging to the classification task and calculate GenRew according to Equation~(\ref{equ:GenRew}), denoted as GR$_{\rm c}$.
Then, 
we select all subjects belonging to the regression task and compute GenRew, 
denoted as GR$_{\rm r}$.
Finally,
we compute the average of GR$_{\rm c}$ and GR$_{\rm r}$ as the task-dimensional GenRew.
(2)~Repeat the above operations for the remaining 5 dimensions 
(i.e., data form/type, structure, training/prioritization scenarios).

\textbf{Prioritization generalizability}.
CertPri always outperforms all baselines against various dimensions in terms of all average GenRew values.
For instance,
in Figure~\ref{fig:generalizabilityRadar},
the area of CertPri covers all baselines.
More specifically,
the average GenRew values of CertPri range from 0.9040 to 0.9813  with average improvements of 32.81\%$\sim$238.54\% compared with baselines.
This is mainly because CertPri's calculation only involves gradient derivation based on back propagation, 
which is easy to implement for any DNN. 
Thus,
CertPri can be generalized to various dimensions.

\begin{center}
\fcolorbox{black}{white!20}{\parbox{0.97\linewidth}
    {
        \emph{\textbf{Answer to RQ4}}:
        CertPri is more generic than baselines in all six dimensions with average generalizability improvements of 32.81\%$\sim$238.54\%.
    }
}
\end{center}

\subsection{Guidance (RQ5)}

\begin{center}
\fcolorbox{black}{gray!20}{\parbox{0.97\linewidth}
    {
        Can CertPri \textit{guide} the retraining of DNNs to improve their performance and robustness?
    }
}
\end{center} 

\begin{figure}
    \centering
    \includegraphics[width=\linewidth]{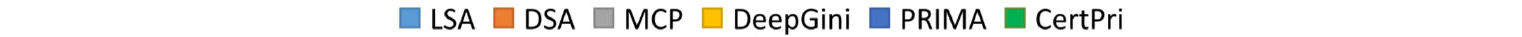} \\
    \subfigure[Accuracy improvement]{
        \includegraphics[width=0.22\linewidth,height=0.32\linewidth]{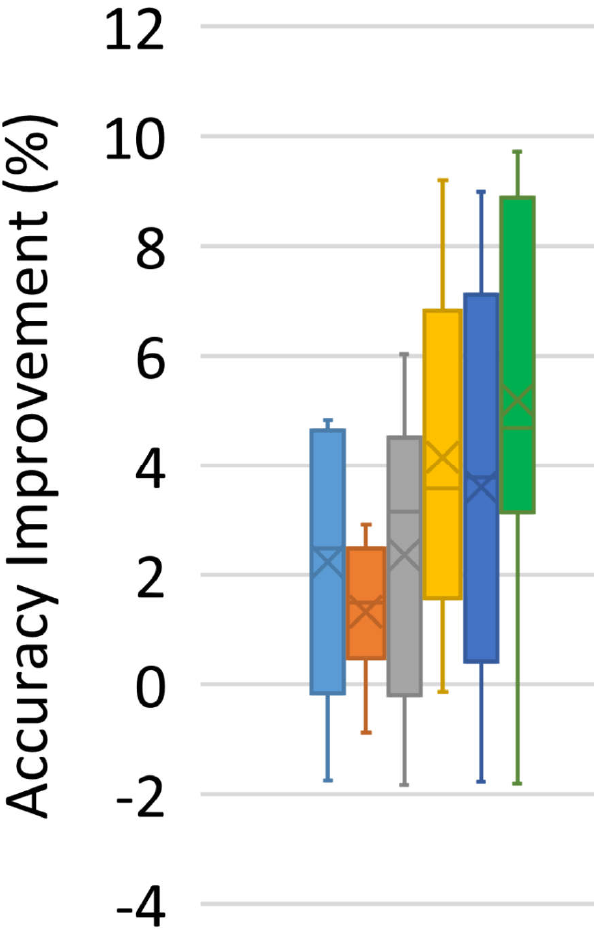}
        \includegraphics[width=0.22\linewidth,height=0.32\linewidth]{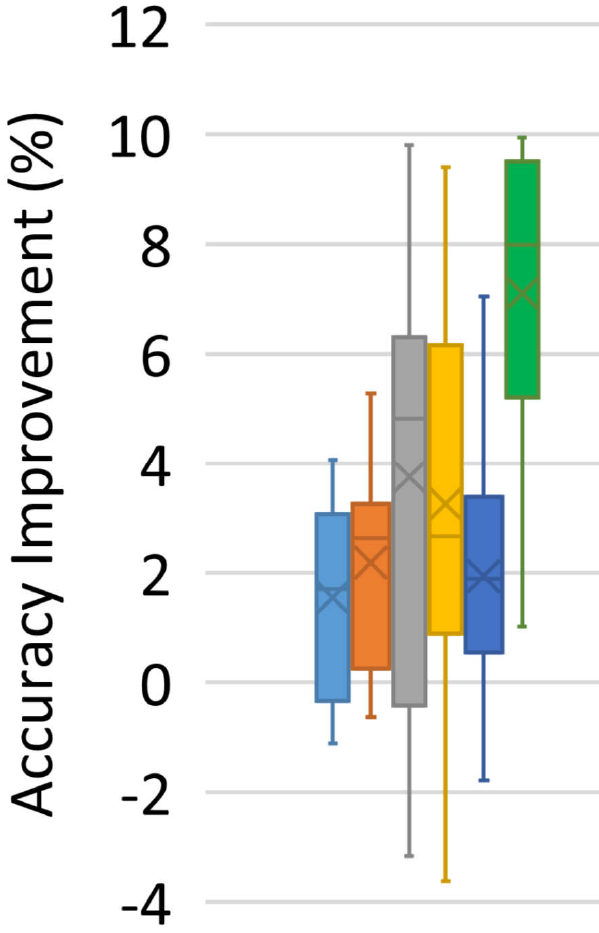}
    } 
    \subfigure[Robustness improvement]{
        \includegraphics[width=0.22\linewidth,height=0.32\linewidth]{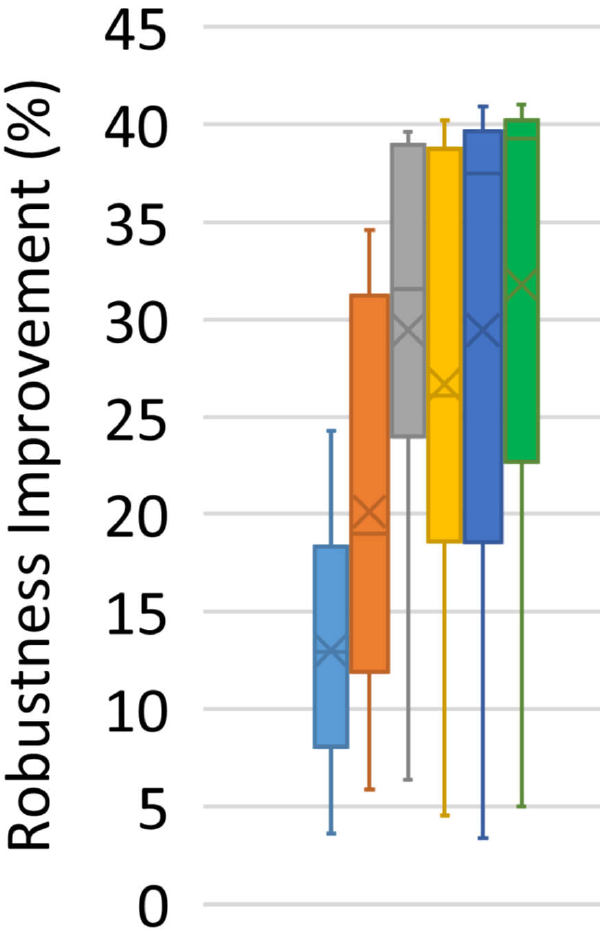}
        \includegraphics[width=0.22\linewidth,height=0.32\linewidth]{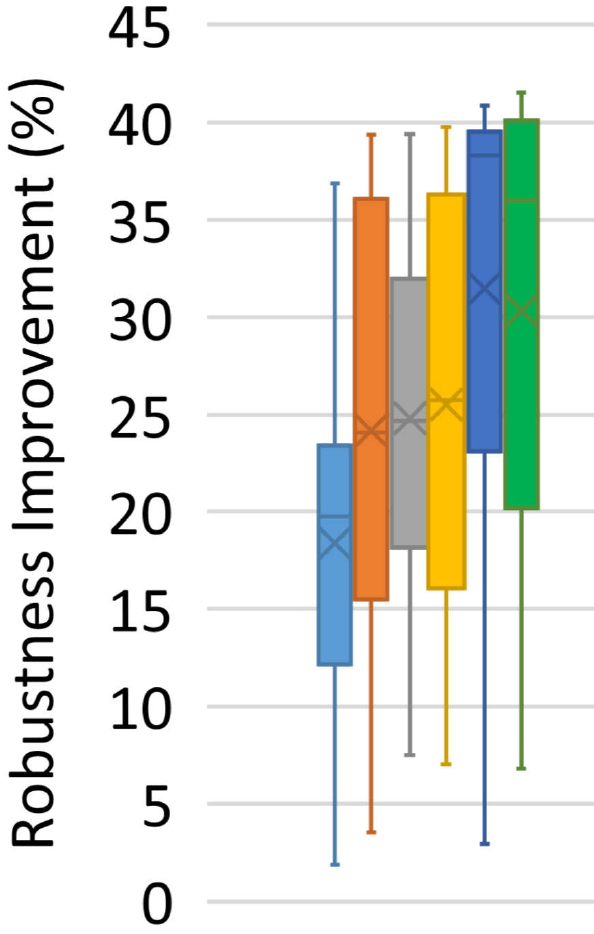}
    }
    \caption{Guidance comparison of accuracy and robustness improvements for different methods under the first 1\% and 5\% prioritized data sampling (from left to right).}
    \label{fig:GuidanceImprove}
\end{figure}

When reporting the guidance, 
we focus on two aspects:
accuracy improvement and robustness improvement for DNNs.
The evaluation results are illustrated as boxplots in Figure~\ref{fig:GuidanceImprove}.

Implementation details for guidance evaluation.
Take the classification on CIFAR10 as an example.
(1)~In terms of performance (ID: 1, 9), 
we sample original data prioritized at the front 1\% and 5\% in the training set. 
We set epoch=5 for retraining due to the small number of data. 
We compare the test accuracy.
(2)~In terms of robustness (ID: 3-5, 11-13), 
we sample adversarial data prioritized at the front 1\% and 5\% in the adversarial set.
We mix the sampled data with the original training set, 
and set epoch=2 for retraining due to a large number of data. 
Repeat the above operations 5 times.

\textbf{Accuracy improvement for DNNs}.
The original test inputs prioritized at the front facilitate model accuracy through retraining,
where CertPri demonstrates SOTA accuracy improvement.
For instance,
in Figure~\ref{fig:GuidanceImprove} (a),
the box position of CertPri is significantly higher than that of baselines.
More specifically,
the average accuracy improvements of CertPri range from 4.50\% to 8.36\%,
which is 1.65$\sim$3.47 times that of baselines.
It demonstrates CertPri's outstanding guidance for the accuracy improvement of DNNs.

\textbf{Robustness improvement for DNNs}.
The adversarial test inputs prioritized at the front facilitate model robustness through retraining,
where CertPri outperforms surprise-based methods and shows competitive performance with confidence-based and mutation-based methods.
For instance,
in Figure~\ref{fig:GuidanceImprove} (b),
the box position of CertPri is significantly higher than that of LSA and DSA,
while close to that of MCP, DeepGini and PRIMA.
It demonstrates CertPri's guidance for the robustness improvement of DNNs.

\begin{center}
\fcolorbox{black}{white!20}{\parbox{0.97\linewidth}
    {
        \emph{\textbf{Answer to RQ5}}:
        CertPri guides DNNs' retraining with only first 1\% or 5\% prioritized test inputs,
        which improves accuracy by 6.15\% and robustness by 31.07\% on average.
    }
}
\end{center}

\section{Threats to Validity\label{Threats}}

Three aspects may become the threats to validity of CertPri.

The \textit{internal} threat to validity mainly lies in the center position determination.
There are differences in movement costs to reach different centers,
especially for non-uniformly distributed data in feature space.
To reduce the internal threat,
we customize the center position for each test input in \textbf{Definition 1}, 
i.e., relative center.
Besides,
we collect a large number of subjects with great diversity,
and perform extensive experiments to verify CertPri’s utility.

The \textit{external} threats to validity mainly lie in 
the non-differentiable component and gradient vanishing.
To reduce these threats,
we further extend \textbf{Theorems 1} and \textbf{2} to a special case of non-differentiable functions,
i.e., a model with ReLU activation,
as shown in \url{https://anonymous.4open.science/r/CertPri/SupplementaryMaterials.pdf}.
Then,
we leverage batch normalization and non-saturating activation to reduce the probability of gradient vanishing and enlarge the sampling radius of the hyper-ball.

The \textit{construct} threats to validity mainly lie in the hyperparameters in CertPri,
including $N_{b}$, $N_{rsb}$ and $\mathcal{R}$ values in \textbf{Algorithms 1}.
Larger hyperparameter values produce better effectiveness, but reduce efficiency.
To reduce the threat from the hyperparameters,
we conduct a double-minimum strategy.
Besides,
the norm value $p$=2 and the extreme value distribution type is Weibull in our experiments.
In future work, 
we can explore prioritization results for various norm types and extreme value distributions.

\section{Related Works\label{RWs}}

To solve the labeling-cost problem in DNN testing,
several works on test input prioritization are proposed~\cite{Feng2020DeepGini,Wang2021PRIMA,Kim2019Guiding,Shen2020MCP,Byun2019DNNsentiment,Zhang2019NSA,Zhang2020NAF,ma2021test,Wang2021PRIMA,Xie2022Boosting}.

From the perspective of statistical analysis, 
there are coverage-based~\cite{Ma2018Deepgauge,Pei2017DeepXplore,Wicker2018Feature,Xie2022NPC}, 
surprise-based~\cite{Zhang2020NAF,Byun2019DNNsentiment,Kim2019Guiding,ma2021test}, 
and confidence-based methods~\cite{Feng2020DeepGini,Shen2020MCP,Zhang2019NSA}.
Feng \textit{et al.}~\cite{Feng2020DeepGini} comprehensively analyzed coverage-based methods and 
concluded that their effectiveness and efficiency are unsatisfactory for prioritization tasks.
To improve effectiveness,
Byun \textit{et al.}~\cite{Byun2019DNNsentiment} prioritized test inputs based on surprise adequacy metrics.
Zhang \textit{et al.}~\cite{Zhang2020NAF} observed the activation pattern of neurons, 
and produced prioritization results according to the activation patterns between the training set and test inputs.
Ma \textit{et al.}~\cite{ma2021test} considered the interaction between test inputs and model uncertainty, 
and determined bug-revealing inputs with higher uncertainty.
These surprise-based methods improve performance, 
but their prioritization results are related to the training set quality.
Furthermore,
Zhang \textit{et al.}~\cite{Zhang2019NSA} prioritized test inputs based on noise sensitivity analysis,
independent of the training set.
Shen \textit{et al.}~\cite{Shen2020MCP} proposed MCP, 
in which clusters test inputs into the boundary areas and specify the priority to select them evenly.
Feng \textit{et al.}~\cite{Feng2020DeepGini} proposed DeepGini, 
which prioritizes test input by measuring set impurity.
These confidence-based methods are effective and efficient, 
but only for classification tasks.

Drawing lessons from the mutation view in software engineering~\cite{lou2015mutation, shin2019empirical}, 
Wang \textit{et al.}~\cite{Wang2021PRIMA} proposed PRIMA, 
which gives priority to test inputs that generate different predictions through diversity mutations 
(i.e., input-level and model-level).
PRIMA demonstrates SOTA performance, 
but cannot be applied to black-box scenarios.

Different from them, 
CertPri prioritizes test inputs based on movement view in feature space.
Besides,
it takes robustness certification into account.
Robustness certification~\cite{Du2021CertRNN,Wei2018CLEVER,ko2019popqorn} provides formal guarantees for DNNs against norm bounded attacks,
which also facilitates CertPri against adaptive attacks.
To the best of our knowledge, 
CertPri is the first to consider the prioritization robustness and introduce formal guarantees to provide certifiability.

\section{Conclusions\label{Conclusions}}

We propose a certifiable prioritization method for bug-revealing test input identification earlier,
CertPri,
to efficiently solve the labeling-cost problem in DNN testing and build trustworthy deep learning systems.
CertPri provides a new perspective on prioritization,
which reduces the problem of measuring misbehavior probability to the problem of measuring the movement difficulty in feature space.
Based on this view,
we give formal guarantees about lower bounds $\gamma_{L}$ on movement cost, 
and compute $\gamma_{L}$ value based on GEVT. 
The priority of each test input is determined in ascending order of $\gamma_{L}$ value. 
Furthermore, we generalize CertPri in black-box scenarios by gradient estimation.
CertPri is compared with baseline on various tasks, data forms, data types, model structures, training and prioritization scenarios.
The results show that CertPri outperforms baselines in terms of effectiveness, efficiency, robustness, generalizability and guidance.

\end{document}